\documentclass[iop]{emulateapj}  

\usepackage{color}
\usepackage{hyperref}  
\usepackage{breakurl}  
\usepackage{xfrac}

\newcommand\kms{\ifmmode{\rm km\thinspace s^{-1}}\else km\thinspace s$^{-1}$\fi}
\newcommand\ms{\ifmmode{\rm m\thinspace s^{-1}}\else m\thinspace s$^{-1}$\fi}
\newcommand\hstar{HII-2147}

\newcommand{\actaa}{Acta Astron.}

\shortauthors{Torres et al.}
\shorttitle{HII-2147}

\begin{document} 

\title{Dynamical masses for the Pleiades binary system HII-2147}

\author{
Guillermo Torres\altaffilmark{1},
Carl Melis\altaffilmark{2},
Adam L.\ Kraus\altaffilmark{3},
Trent J.\ Dupuy\altaffilmark{4},
Jeffrey K.\ Chilcote\altaffilmark{5}, and \\
Justin R.\ Crepp\altaffilmark{5}
}

\altaffiltext{1}{Center for Astrophysics $\vert$ Harvard \&
  Smithsonian, Cambridge, MA 02138, USA; gtorres@cfa.harvard.edu}

\altaffiltext{2}{Center for Astrophysics and Space Sciences, Univ.\ of
  California at San Diego, La Jolla, CA 92093, USA}

\altaffiltext{3}{Department of Astronomy, The Univ.\ of Texas at
  Austin, Austin, TX 78712, USA}

\altaffiltext{4}{Gemini Observatory, Northern Operations Center, Hilo,
  HI 96720, USA}

\altaffiltext{5}{Department of Physics, Univ.\ of Notre Dame, Notre
  Dame, IN 46556, USA}

\begin{abstract}

We report our long-term spectroscopic monitoring of the Pleiades
member \hstar, which has previously been spatially resolved at radio
wavelengths in VLBI observations. It has also been claimed to be a
(presumably short-period) double-lined spectroscopic binary with
relatively sharp lines, although no orbit has ever been published.
Examination of our new spectroscopic material, and of the historical
radial velocities, shows that the current and previous spectra are
best interpreted as showing only a single set of lines of a moderately
rapidly rotating star with slowly variable radial velocity, which is
one of the sources detected by VLBI. We combine our own and other
velocities with the VLBI measurements and new adaptive optics
observations to derive the first astrometric-spectroscopic orbit of
the G5\,+\,G9 pair, with a period of $18.18 \pm 0.11$ years. We infer
dynamical masses of $0.897 \pm 0.022~M_{\sun}$ for the
spectroscopically visible star and $0.978 \pm 0.024~M_{\sun}$ for the
other, along with a distance of $136.78^{+0.50}_{-0.46}$~pc. The lack
of detection of the lines of the more massive component in our spectra
can be adequately explained if it is rotating much more rapidly than
the star we see. This is consistent with the observation that the
lines of the secondary are shallower than expected for a star of its
spectral type.

\end{abstract}

\section{Introduction}
\label{sec:introduction}

The Pleiades cluster has served as a valuable laboratory for stellar
astrophysics for decades. Astrometric and spectroscopic surveys have
found many binary and multiple systems among its $\sim$1500 members
\citep[e.g.,][]{Rosvick:1992, Mermilliod:1992, Mermilliod:1997,
  Bouvier:1997, Hillenbrand:2018}, and yet very few have had their
component masses --- the most basic stellar property --- determined
reliably.  To our knowledge there are only three examples: the
interferometric-spectroscopic binary Atlas \citep[27\thinspace Tau, HD
  23850;][]{Zwahlen:2004}, the eclipsing system HD\thinspace 23642
\citep[V1229\thinspace Tau;][]{Torres:2003, Munari:2004,
  Southworth:2005, Groenewegen:2007, David:2016}, and more recently
HCG\thinspace 76 \citep[V612\thinspace Tau;][]{David:2016}, also an
eclipsing binary with low-mass components.

Two other eclipsing systems have been found in the Pleiades in recent
years that may also eventually lead to accurate dynamical mass
determinations.  One is HII-2407 \citep[V1283\thinspace
  Tau;][]{David:2015, David:2016}, which is so far only a single-lined
spectroscopic binary and must therefore await detection of the
secondary lines before the masses can be determined without
assumptions. The other is MHO\thinspace 9 \citep[EPIC
  211075914;][]{David:2016}, a long period (42.8~days), low-mass, and
very faint system ($V = 19.02$) that is double-lined but for which the
available data are still insufficient to obtain meaningful estimates
of its properties.

In this paper we report our long-term spectroscopic monitoring of the
Pleiades member \hstar\ (V1282\thinspace Tau, 2MASS\thinspace
J03490610+2346525, $V = 10.80$, $B-V = +0.83$). The second data
release (DR2) of the {\it Gaia\/} catalog \citep{Gaia:2018} reports a
trigonometric parallax of $7.209 \pm 0.051$~mas, corresponding to a
distance of $138.7 \pm 1.0$~pc. The object has been spatially resolved
at radio wavelengths into a $\sim$60~mas binary with the technique of
very long baseline interferometry \citep[VLBI;][]{Melis:2014}, in the
course of a program prior to {\it Gaia} to determine trigonometric
parallaxes in the cluster. As is the case for many stars in the
Pleiades, \hstar\ is chromospherically active and displays photometric
variability likely caused by spots on its surface, and on this basis
it has been listed as a member of the RS~CVn class. It is also an
X-ray source \citep{Voges:1999, Freund:2018} with flaring activity
\citep{Gagne:1995}.  By combining our spectroscopy with new imaging
observations and other existing radial-velocity measurements from the
literature, we are able to determine the masses of the components,
making it the fourth system in the cluster with such empirical
measurements.

We have organized the paper as follows. Section~\ref{sec:spectroscopy}
discusses the historical radial-velocity measurements of \hstar\ on
the basis of which it was claimed to be a double-lined spectroscopic
binary, but for which an orbit was mysteriously never derived. After
showing that interpretation of the system to be incorrect, we present
in the same section our own spectroscopic monitoring spanning more
than 37 years. The VLBI observations are described in
Section~\ref{sec:vlbi}, and measurements from new adaptive optics
imaging that resolve the pair are presented in
Section~\ref{sec:ao}. Then in Section~\ref{sec:analysis} we analyze
all of the observations together, including brightness measurements,
to derive the first orbital solution for \hstar\ and infer the
component masses. Alternate scenarios are presented here as well, to
explain the lack of detection of the lines of the primary star in our
spectra. In Section~\ref{sec:rotation} we review the measurements of
the rotation period and discuss their implications. We conclude in
Section~\ref{sec:discussion} with a discussion of the results.

\section{Spectroscopy}
\label{sec:spectroscopy}

\subsection{CORAVEL observations}
\label{sec:coravel}

In their spectroscopic investigation of the Pleiades cluster,
\cite{Mermilliod:1992} reported \hstar\ to be a double-lined
spectroscopic binary based on observations with the CORAVEL
radial-velocity scanner \citep{Baranne:1979} gathered between 1978 and
1991. They indicated the object has a broad cross-correlation profile,
and claimed that if taken at face value it would imply a spin rate
that is too high for a star of the assumed spectral type (K0V).  In
support of their conclusion that it is double-lined, they pointed out
the location of \hstar\ above the main sequence in the color-magnitude
diagram of the cluster. While they were unable to obtain an orbital
solution from the measured velocities despite having some 50
observations, they noted that if the observations are reduced as if
the star were single-lined, they show a long-term trend indicative of
a period longer than 6000 days. A subsequent paper by
\cite{Queloz:1998} reported individual $v \sin i$ measurements for the
``primary'' and ``secondary'' of $6.9 \pm 3.2$ and $10.8 \pm
2.3$~\kms, respectively, based on the same CORAVEL observations.

\begin{figure}
\epsscale{1.15}
\plotone{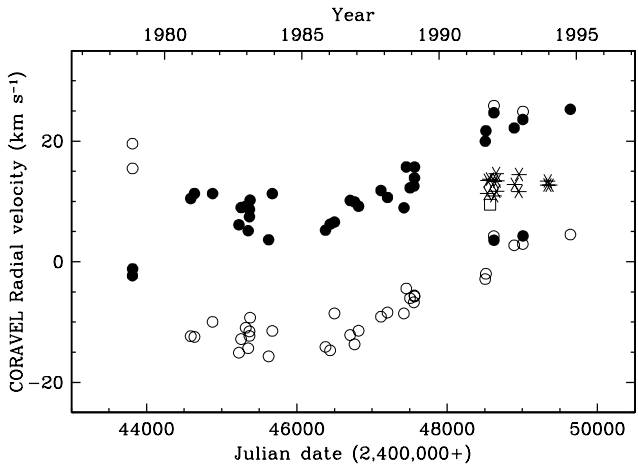}

\figcaption{CORAVEL radial-velocity measurements of \hstar\ as
  published by \cite{Mermilliod:2009}. The filled and open circles
  correspond, respectively, to the ``primary'' and ``secondary''
  component assignments, as given by those authors. Star symbols
  represent single measurements that they attributed to the
  ``primary'', and the square symbol is a measurement with no
  component assignation.\label{fig:coravel}}

\end{figure}

The individual CORAVEL radial velocities for the ``primary'' and
``secondary'', augmented with new observations with the same
instrument between 1991 and 1994, were published by
\cite{Mermilliod:2009} and are shown in Figure~\ref{fig:coravel}.
Occasionally only a single velocity was measured, and was attributed
to the ``primary'' component. The long-term trend is certainly
obvious, but we are skeptical that the evidence supports the claim
that the visible object is double-lined (presumably with a short
orbital period).\footnote{Interestingly, \cite{Mermilliod:2009} appear
  less certain about their original claim, listing the object as
  ``SB2?''  and reporting the projected rotational velocity as $v \sin
  i = 27.1 \pm 2.7~\kms$. Furthermore, the summary properties in their
  Table~11 give the scatter in the velocities of \hstar\ as 5.13~\kms,
  which seems inconsistent with the spread in the individual
  velocities as published, unless the value refers to only one
  component, or perhaps to the average of the
  two.\label{foot:mermilliod}} Instead, we contend there is a single
broad cross-correlation profile, and that the ``primary'' and
``secondary'' velocities measured by \cite{Mermilliod:1992} are
spurious, resulting from the interpretation of the profile as the sum
of two widely separated, narrower profiles. We base this contention on
the following. \emph{i)} The ``primary''/``secondary'' velocity
separation seems to always be about 20~\kms, whereas for a (presumably
short-period) double-lined binary observed at random times one would
expect to see some range. The only measurements that depart from this
pattern are ones in which only a single velocity was measured, and
those happen to fall very nearly at the average of the long term
trends followed by the ``primary'' and ``secondary'' measurements.
\emph{ii)} Experiments with our own observations, described below,
where we treated the single broad cross-correlation peak as if it were
due to two stars, also gave a nearly constant velocity separation of
20~\kms, and no convincing evidence of orbital motion. \emph{iii)} As
a more direct indication of the same effect, the measured width of the
cross-correlation profiles of our own spectra show no significant
change with time, which one would expect to see if the broadened
profile were the result of two narrower ones moving relative to one
another.  \emph{iv)} The \cite{Mermilliod:1992} argument that a single
broad correlation profile implies an excessive spin rate was likely
based on a typical rotational period for a K star in the Pleiades
(roughly 6 days) that has slowed down and settled on the rotational
sequence. However, there is in fact a subsample of much more rapidly
rotating cluster members of the same spectral type
\citep[e.g.,][]{Hartman:2010, Rebull:2016} known as ultra-fast
rotators (UFRs), whose origin is still being debated
\citep[see][]{Butler:1987, Barnes:1996, Brown:2014, Garraffo:2018}. As
mentioned earlier, \hstar\ is an active and spotted star. Several
direct measurements of its rotation period from the spot modulations
now place it at well under a day (see Section~\ref{sec:rotation}),
making it a member of the class of UFRs; \emph{v)} The excess
brightness of \hstar\ in the color-magnitude diagram can be adequately
explained by the companion in the long-term orbit, as we will show
later.

We therefore proceed under the assumption that the CORAVEL
observations recorded a single, broad set of lines corresponding to a
rapidly rotating star, and for the purposes of the orbital analysis
below we will approximate the centroid of those broad correlation
profiles by the straight average of the ``primary'' and ``secondary''
velocities as reported by \cite{Mermilliod:2009}.  We will also adopt
provisional uncertainties given by the quadrature sum of the
individual ``primary'' and ``secondary'' velocity errors as published.

While Figure~\ref{fig:coravel} clearly demonstrates that the star is
participating in a long-period binary orbit, neither the CORAVEL
observations nor our own (see below) show direct evidence of the
spectral lines of the companion. In principle, it is therefore
possible that the broad profile we see is the result of the blending
(flux-weighted average) of the lines of those two objects, especially
given that the long orbital period would imply relatively small
radial-velocity amplitudes that could prevent the detection of two
separate sets of lines.  In that case, one would expect changes in the
width of the correlation profile with time as the velocity separation
widens and narrows throughout the orbit.  However, no such changes are
seen, as mentioned above, which argues that the correlation profile is
dominated by the lines of only one object.

\subsection{New spectroscopy}
\label{sec:newspectroscopy}

Spectroscopic monitoring of \hstar\ at the Center for Astrophysics
(CfA) began in 1982, and continued until 2020 with several different
instruments and telescopes, as follows. The initial observation was
made with CfA Digital Speedometer \citep[DS;][]{Latham:1992} on the
4.5m-equivalent Multiple Mirror Telescope (Mount Hopkins, Arizona,
USA) before its conversion to a monolithic 6.5m telescope. Five
additional observations through 2003 December were made with copies of
this instrument on the 1.5m Tillinghast Reflector Echelle Spectrograph
at the Fred L.\ Whipple Observatory (also on Mount Hopkins) and with
the (now closed) 1.5m Wyeth reflector at the Oak Ridge Observatory
(Harvard, Massachusetts, USA). These instruments were equipped with
intensified photon-counting Reticon detectors, and recorded a single
echelle order 45\thinspace \AA\ wide centered at a wavelength near
5187\thinspace \AA\ containing the \ion{Mg}{1}~b triplet, at a
resolving power of $R \approx 35,\!000$. Reductions were performed
with a custom pipeline, and the wavelength calibration was based on
exposures of a thorium-argon (ThAr) lamp before and after each science
exposure. Twilight observations were obtained regularly at dusk and
dawn and used to maintain the velocity zero point
\citep[see][]{Latham:1992} by applying small run-to-run corrections
generally smaller than 2~\kms.  The signal-to-noise ratios of these
six observations range between about 10 and 24 per resolution element
of 8.5~\kms.

Beginning in 2011, the observations were continued with the
Tillinghast Reflector Echelle Spectrograph
\citep[TRES;][]{Szentgyorgyi:2007, Furesz:2008}, a bench-mounted
fiber-fed instrument providing a resolving power of $R \approx
44,\!000$ and covering the wavelength region 3800--9100\thinspace
\AA\ in 51 orders. For the order centered at $\sim$5187\thinspace
\AA\ that we used for the velocity determinations, the signal-to-noise
ratios range from about 15 to 48 per resolution element of 6.8~\kms. A
total of 35 spectra of \hstar\ were collected with this instrument
through 2020 March. Wavelength calibrations relied on ThAr exposures
preceding and following the science frames, as above, and the
reductions were carried out with a separate dedicated pipeline. The
velocity zero point was monitored by taking exposures of several IAU
radial-velocity standard stars, and typically varies by less than
50~\ms, which is more than sufficient for this work. Spectra of the
same standard stars from the DS were used to place both sets of
observations on a common velocity system, which is within about
0.14~\kms\ of the reference frame defined by observations of minor
planets in the solar system \citep[see][]{Stefanik:1999, Latham:2002}.

\begin{figure}
\epsscale{1.15}
\plotone{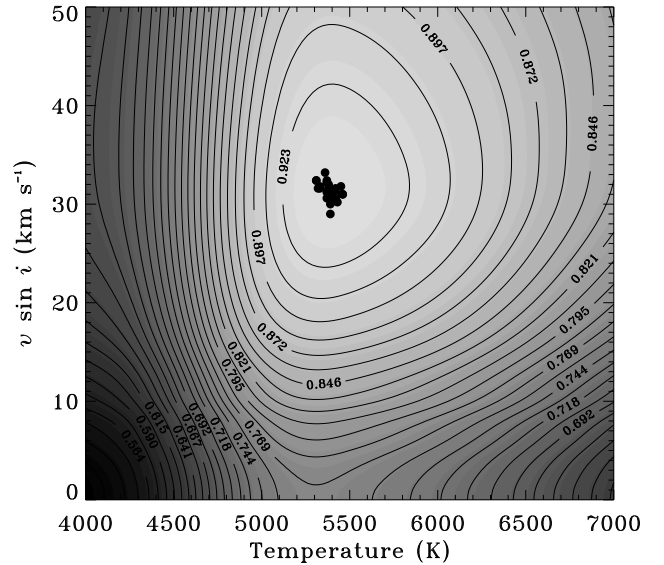}

\figcaption{Determination of the temperature and rotational broadening
  of \hstar. The contours correspond to equal values of the
  cross-correlation function averaged over all exposures, and the
  points mark the results for individual
  spectra.\label{fig:templategrid}}

\end{figure}

All our spectra appear single-lined, and show significant rotational
broadening. Radial velocities were measured by cross-correlation using
the IRAF task {\tt xcsao}.\footnote{IRAF is distributed by the
  National Optical Astronomy Observatories, which is operated by the
  Association of Universities for Research in Astronomy, Inc., under
  contract with the National Science Foundation.} The template was
selected from a pre-computed library of synthetic spectra based on
model atmospheres by R.\ L.\ Kurucz, and a line list tuned to better
match the spectra of real stars \citep[see][]{Nordstrom:1994,
  Latham:2002}.  The wavelength region covered by these templates is
limited to $\sim$300\thinspace \AA\ centered at 5187\thinspace \AA,
and the two most important parameters for velocity determinations are
the effective temperature ($T_{\rm eff}$) and rotational broadening
($v \sin i$).  The optimal values of these parameters were determined
by running a grid of cross correlations over broad ranges for a fixed
surface gravity of $\log g = 4.5$ and solar metallicity, and adopting
the $T_{\rm eff}$ and $v \sin i$ values giving the highest correlation
averaged over all exposures, following \cite{Torres:2002}. For this
analysis we used the more numerous and higher quality TRES spectra.
The result may be visualized in Figure~\ref{fig:templategrid}, and
yielded a best temperature of $5390 \pm 100$~K and $v \sin i = 31 \pm
2~\kms$.  The temperature corresponds roughly to spectral type G9,
based on the tabulations of \cite{Gray:1992} or \cite{Pecaut:2013}.
For the radial velocity determinations we adopted template parameters
of 5500~K and 30~\kms, the nearest in our grid.  The heliocentric
radial velocities along with their uncertainties are listed in
Table~\ref{tab:cfarvs}. Given that \hstar\ is chromospherically
active, these uncertainties likely include a contribution from
distortions in the line profiles caused by spots moving across the
stellar disk.

\setlength{\tabcolsep}{9pt}  
\begin{deluxetable}{lcc}
\tablewidth{0pc}
\tablecaption{Heliocentric Radial-velocity Measurements of
  \hstar\ from CfA \label{tab:cfarvs}}
\tablehead{
\colhead{HJD} &
\colhead{$RV$} &
\colhead{$\sigma_{\rm RV}$}
\\
\colhead{(2,400,000$+$)} &
\colhead{(\kms)} &
\colhead{(\kms)}
}
\startdata

45302.8166 &   3.50   &  1.50 \\ 
45308.6940 &   0.61   &  2.57 \\
45341.7564 &  $-$3.82\phs &  2.46 \\
51238.7338 &   4.29   &  0.90 \\
52920.8970 &  $-$3.23\phs &  3.13 \\
52975.7620 &  $-$4.05\phs &  1.81 \\
55846.9010 &  12.15\phn  &   1.18 \\ 
55964.7124 &  12.86\phn  &   1.80 \\ 
56322.7765 &   9.42  &   1.53 \\ 
56347.6118 &  11.15\phn  &   1.03 \\
56617.6476 &   9.61  &   0.97 \\ 
56644.7420 &   9.83  &   0.92 \\ 
56669.6778 &   8.89  &   0.71 \\
56675.7899 &   9.39  &   0.74 \\
56693.6257 &   9.79  &   0.88 \\ 
56694.6127 &  10.95\phn  &   0.91 \\ 
56731.6706 &   8.59  &   0.78 \\
56902.9993 &   8.35  &   0.70 \\ 
56927.0248 &   8.62  &   0.97 \\ 
56934.8492 &   8.55  &   0.78 \\ 
56944.8326 &   7.87  &   0.72 \\ 
56959.9076 &   9.10  &   0.73 \\
56962.7465 &   8.45  &   0.76 \\
56970.8194 &   7.82  &   0.69 \\
56972.7779 &   9.21  &   0.68 \\
56978.7084 &   9.34  &   0.75 \\ 
56987.6674 &   9.70  &   0.73 \\
57001.7034 &   8.52  &   0.64 \\
57026.7480 &   8.69  &   0.60 \\
57062.6000 &   8.41  &   0.89 \\
57090.6127 &   8.44  &   0.63 \\
57294.0219 &   7.12  &   0.72 \\ 
58157.5983 &   3.58  &   0.68 \\
58424.7888 &   0.89  &   1.04 \\
58451.7360 &   1.27  &   1.11 \\ 
58476.7761 &   0.68  &   0.79 \\
58503.7663 &  $-$0.19\phs  &   0.66 \\
58770.8336 &  $-$0.88\phs  &   1.32 \\ 
58886.7129 &   0.08  &   1.20 \\
58923.6722 &   0.74  &   1.67 \\
58924.6516 &   0.17  &   0.68 \\
47467.035  &   7.4\phn   &   1.30 \\ 
52536.     &  $-$4.72\phs  &   0.75     
\enddata

\tablecomments{The last two measurements are from
  \cite{Soderblom:1993} and \cite{White:2007}, adjusted by +0.10 and
  $-0.60~\kms$, respectively, to place them on the same zero point as
  the CfA velocities (see text).}

\end{deluxetable}
\setlength{\tabcolsep}{6pt}  

The rotational broadening we measure for \hstar\ is consistent with
other estimates from the literature: \cite{Soderblom:1993} reported $v
\sin i = 27 \pm 3~\kms$, \cite{White:2007} reported $38.34 \pm
2.32~\kms$ (although they pointed out that their result is probably
overestimated by 6--9~\kms), and \cite{Mermilliod:2009} gave $27.1 \pm
2.7~\kms$.  Several of our spectra show signs of contamination from
moonlight. To prevent this from biasing the measured velocities, we
reanalyzed the affected spectra with the two-dimensional
cross-correlation algorithm TODCOR \citep{Zucker:1994}, which uses two
templates, one for each set of spectral lines. We adopted the same
template as above for \hstar, and a solar template for the
contaminating light. The final heliocentric radial velocities after
these corrections, along with their uncertainties, are listed in
Table~\ref{tab:cfarvs}.

In order to test the idea that the broad cross-correlation profile of
\hstar\ may be the sum of two narrower profiles, as claimed by
\cite{Mermilliod:1992}, we again used TODCOR with identical
temperatures as before for the ``primary'' and ``secondary''
templates, and rotational broadenings of 6 and 10~\kms, near those
reported by \cite{Queloz:1998}. The resulting ``primary'' and
``secondary'' velocities showed a nearly constant separation of about
20~\kms, as noted earlier of the CORAVEL velocities. After removal of
an obvious long-term trend, we were unable to establish a reasonable
double-lined orbital solution, as had been found earlier by
\cite{Mermilliod:1992} for the CORAVEL measurements. We therefore take
this as evidence that our spectroscopic observations are best
interpreted as featuring a single broad cross-correlation profile.

Additional spectroscopic observations were secured at the Lick
Observatory with the Shane 3m telescope.  Light was fed into the
coud\'{e} focus that houses the Hamilton echelle spectrograph
\citep{Vogt:1987}, and all observations used Dewar\#4 with the
detector windowed to record light from 3850 to 9250\thinspace
\AA. With the exception of the Nov.\ 2012 epoch (HJD 2,456,257.87), a
640\,$\mu$m-wide slit was employed yielding a spectral resolving power
of $\sim$62,000, as measured from the FWHM of single titanium-argon
(TiAr) arc lines in comparison spectra. For the Nov.\ 2012 epoch, an
800\,$\mu$m-wide slit was used resulting in $R \approx 40,000$.  Data
reduction for the Hamilton echelle spectrograph with IRAF tasks is
outlined in detail in Lick Technical Report
No.\ 74.\footnote{\url{http://astronomy.nmsu.edu/cwc/Software/irafman/manual.html}}
Briefly, the data were bias subtracted, flat-fielded, extracted, and
finally wavelength-calibrated with TiAr arc lamp spectra \citep[see][]
{Pakhomov:2013}. Signal-to-noise ratios for these observations range
from 65 to 100 per pixel at a mean wavelength of 6000\thinspace \AA. A
stable radial velocity standard star \citep[usually HR\thinspace 124
  or HD\thinspace 4203;][]{Nidever:2002} was observed each night along
with \hstar, and used as the template for the cross-correlations, with
adopted absolute velocities as reported by \cite{Nidever:2002}.  The 9
Lick radial-velocity measurements and uncertainties, based typically
on the five best spectral orders, are given in
Table~\ref{tab:hamilton} in the heliocentric frame.

\setlength{\tabcolsep}{9pt}  
\begin{deluxetable}{lcc}
\tablewidth{0pc}
\tablecaption{Heliocentric Radial-velocity Measurements of
  \hstar\ from Lick \label{tab:hamilton}}
\tablehead{
\colhead{HJD} &
\colhead{$RV$} &
\colhead{$\sigma_{\rm RV}$}
\\
\colhead{(2,400,000$+$)} &
\colhead{(\kms)} &
\colhead{(\kms)}
}
\startdata
56257.86531  &  11.8\phn  &   0.5  \\
56581.89577  &  9.4  &   0.4  \\
56610.95804  &  11.2\phn  &   0.6  \\
56678.80577  &  9.9  &   0.5  \\
56711.72436  &  10.3\phn  &   0.6  \\
56904.95041  &  8.2 &  0.5 \\
57262.94947  &   7.9  &   0.5  \\
57349.81417  &   5.0  &   0.6  \\
57752.75111  &   5.4  &   0.6
\enddata

\end{deluxetable}
\setlength{\tabcolsep}{6pt}  

\subsection{Archival radial velocities}
\label{sec:archival}

Aside from the CORAVEL and our own observations, only a few isolated
measurements of the radial velocity of \hstar\ have appeared in the
literature. \cite{Soderblom:1993} published a single measurement from
1988, and \cite{White:2007} published another from 2002. The latter
happens to fall at a time near a radial-velocity minimum (see below),
when few other observations are available, making it potentially
constraining. In order to place those two measurements on the same
reference frame as ours, we compared the velocities of other Pleiades
stars measured by these authors with observations of the same stars
from our own survey of the cluster.  Based on 7 stars in common with
\cite{Soderblom:1993}, we established a correction of $+0.10~\kms$ to
bring their measurement of \hstar\ onto the CfA zero point. Similarly,
from 10 stars we determined a correction of $-0.60~\kms$ for the
\cite{White:2007} measurement. The adjusted velocities from these two
literature sources are included at the end of Table~\ref{tab:cfarvs}.
Additionally, \hstar\ is included in the {\it Gaia}/DR2 catalog
\citep{Gaia:2018} with the identifier number 66503449709270400. The
average radial velocity reported there is $6.1 \pm 6.2~\kms$, obtained
from 7 transits over a period of 22 months.  The large uncertainty is
due perhaps to real changes, or to reduced precision because of the
rotational broadening of the spectral lines. However, as it is only an
average and the velocity is changing, we have chosen not to make use
of that value.

\subsection{Evidence for the companion}
\label{sec:companion}

\begin{figure}
\epsscale{1.15}
\plotone{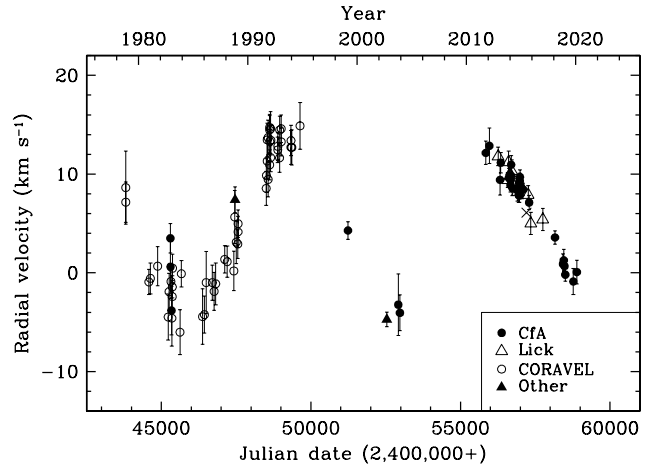}

\figcaption{All available radial velocities for \hstar\ from the
  CORAVEL (average of the \citealt{Mermilliod:2009} ``primary'' and
  ``secondary'' velocities), our own measurements (CfA, Lick), and
  other measurements from the literature. The value from the {\it
    Gaia}/DR2 catalog is also shown as a cross at epoch 2015.5,
  although we do not use it in the analysis that follows. An
  $\sim$18-yr cycle is revealed, which is covered more than
  twice. \label{fig:allrvs}}

\end{figure}

The combination of all available radial velocities reveals a
periodicity of about 18 years that is shown in
Figure~\ref{fig:allrvs}.  If we assume, based on its temperature, that
the star with visible lines is somewhat less massive than the Sun, a
preliminary spectroscopic orbital solution then yields the unexpected
result that the companion is more massive than the star we see.  This
raises the question of why we do not detect the lines of this
companion in our spectra. One possible explanation might be very rapid
rotation. Based on previous experience with similar spectroscopic
material obtained for other objects with the DS and TRES instruments,
we estimate that if the companion were rotating at a projected
velocity of $\sim$100~\kms\ or more, its lines would be very broad and
difficult to distinguish, particularly since they would be heavily
blended with those of the visible star. This possibility seems
consistent with the evidence mentioned earlier of a very short
rotation period ($<$ 1 day) associated with \hstar, based on
photometric modulation due to spots, provided that periodic signal
comes from the companion. Alternatively, the companion may itself be a
close binary, which could make its detection more challenging.

\begin{figure}
\epsscale{1.15}
\plotone{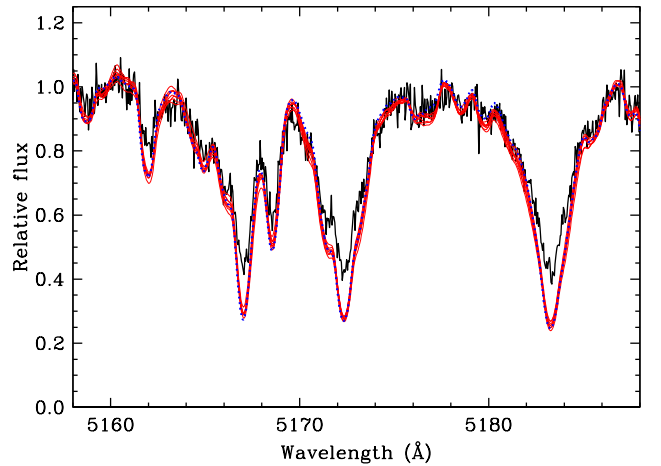}

\figcaption{Portion of one of our TRES spectra of \hstar\ from
  HJD~2,456,902.9993 (noisier black line), compared against observed
  spectra for 8 sharp-lined Pleiades stars of similar temperature
  (thin red lines), broadened to match the $v \sin i$ of \hstar. The
  features of \hstar\ appear weaker, suggesting dilution by the
  continuum of star~A. Also shown is the template we used for the
  radial velocity determinations (dotted blue line), which provides a
  good representation of the spectra of the single stars.
  \label{fig:obsspectra}}

\end{figure}

While the spectral lines of the companion are not seen directly, there
is indirect evidence of its presence from the fact that the features
of the visible star appear shallower than expected for a star of its
spectral type. To illustrate this, we selected TRES spectra of 8
sharp-lined stars from our ongoing spectroscopic survey of the
Pleiades that are within 50\thinspace K of the measured temperature of
\hstar\ (determined in the same way as described in
Section~\ref{sec:newspectroscopy}), and are not known to be binaries.
We broadened these observed spectra using the rotational kernel of
\cite{Gray:1992} to match the rotational broadening of \hstar, and
shifted them to a common wavelength scale. These 8 stars are compared
in Figure~\ref{fig:obsspectra} against an exposure of \hstar, clearly
showing the veiling effect attributable to star~A.  For reference, we
overplot the synthetic template used earlier to derive the velocities
(blue dotted line), which is seen to be a good representation of the
single stars. A more quantitative discussion of the line dilution will
be presented below in Section~\ref{sec:discussion}.  We note, finally,
that a veiling effect for \hstar\ was reported also by
\cite{Kounkel:2019}, based on a near-infrared spectrum in the $H$
band.

\section{VLBI observations}
\label{sec:vlbi}

\hstar\ was observed by \cite{Melis:2014} between 2011 and 2013, as
part of a program to determine trigonometric parallaxes of several
Pleiades members using very long-baseline interferometry (VLBI) at
radio frequencies (a continuum frequency of 8.4~GHz, corresponding to
$\sim$3.6~cm). The original goal was to address the disagreement
between the {\it Hipparcos\/} measurement of the distance to the
cluster and determinations by other methods.  \hstar\ was found to be
a double source with a separation around 60~mas, and accurate
positions for the equinox J2000 referenced to the background quasar
J0347+2339 were reported for both components, oriented NW and SE at
the time. In deriving a parallax for \hstar, the authors included
acceleration terms in right ascension and declination in order to
model the slow motion in an unknown orbit over the two-year observing
period. Orbital motion was in fact evident from the opposite signs of
the acceleration terms for the two components. This astrometric motion
corresponds to the same orbit suggested by the radial velocities,
although the portion covered by the astrometry is small (only
$\sim$10\% of a cycle). Nevertheless, the VLBI observations do help to
constrain that orbit by providing the angular scale, and we use them
below for that purpose under the assumption that the source of the
radio emission is coincident with the center of each stellar disk.
These observations also constrain the parallax of the system. However,
they do not identify which of the components is the one we observed
spectroscopically.

\section{Adaptive optics imaging}
\label{sec:ao}

There are no observations of \hstar\ in the literature that resolve
the pair, although several attempts have been made at both optical and
near-infrared (NIR) wavelengths \citep{Bouvier:1997, Metchev:2009,
  Mason:2009}, including a lunar occultation observation by
\cite{Richichi:2012}.

Due to the limited phase coverage of the VLBI observations, and in
order to supplement the astrometry, adaptive optics (AO) imaging
observations were carried out for this work with the Mauna Kea
Observatory Keck Adaptive Optics system \citep{Wizinowich:2000,
  Wizinowich:2013} on UT 03 and 05 February 2013, as well as on UT 03
November 2019.  The Keck Adaptive Optics system was fed into NIRC2, a
camera with a $1024 \times 1024$ InSb Aladdin-3 array. All NIRC2
observations were performed in the ``Narrow'' camera mode, with a
plate scale of $\sim$0\farcs01\thinspace pixel$^{-1}$.  \hstar\ served
as its own guide star for the AO system.

The observations on the first two nights consisted of a 5-point dither
pattern sequence. On UT 03 February 2013, short exposures
($\leq$2~seconds) in which the binary system was not saturated in each
frame were obtained in each of the $J$ (1.248~$\mu$m), $H$
(1.633~$\mu$m), and $K_{\rm S}$ (2.146~$\mu$m) bands.  High quality
adaptive optics corrections were obtained with average Strehl ratios
of $\sim$0.6 in the $K_{\rm S}$ band, $\sim$0.35 in $H$, and $\sim$0.2
in $J$.  On UT 05 February 2013, longer exposures (50~seconds) were
obtained in the $K_{\rm S}$ band to search for faint companions to the
binary system (no such companions were seen).

Unsaturated observations on UT 03 November 2019 were obtained through
$J$, $H$, and $K^{\prime}$ (2.124~$\mu$m) filters, and were taken with
a 0.2 second integration time using 10 coadds and the CDS sampling
mode. The NIRC2 subarray was set to $512\times512$ pixels.
Observations were gathered in a 3-point dither pattern of five
exposures at each location, with each leg being 1\farcs5, for a total
of 15 observations in each filter. The average differential image
motion monitor (DIMM) measure of the seeing during the observations
was 0\farcs38.  Observations of V1090~Tau were made as a PSF reference
calibrator star. Those exposures were taken with an identical
subarray, integration time, and coadd as the \hstar\ observations.

The data were analyzed using custom scripts that perform the standard
tasks of nonlinearity correction, dark subtraction, and flat fielding.
These scripts also perform ``de-striping'', a rectification of
spatially correlated detector noise that is mirrored across the
quadrants of the NIRC2 detector, and which dominates the photometric
noise budget in the readnoise-limited regime. Finally, the scripts use
bilinear interpolation to estimate values for pixels impacted by
cosmic rays, as well as for the hot pixels and dead pixels that were
identified in super-stacks of darks and flats, as described by
\cite{Kraus:2016}.

Each science frame was then iteratively analyzed with PSF-fitting
photometry to find the best-fit template that minimized the residuals
after PSF subtraction, as described by \cite{Kraus:2016}. The first
stage found the best-fit binary model (separation $\rho$, position
angle (P.A.) $\theta$, and magnitude difference $\Delta m$) given an
empirical template of a single star. The second stage then measured
the $\chi^2$ goodness-of-fit for the 1000 archival frames of single
stars that were taken closest in time and in the same filter, doubling
each potential template with the same binary parameters and then
scaling and subtracting it from the science frame. The two steps were
then repeated until the same empirical template PSF was found to yield
the lowest $\chi^2$ value in two consecutive iterations, and that PSF
was adopted as the template for that science frame.

\setlength{\tabcolsep}{3pt}  
\begin{deluxetable}{lcccc}
\tablewidth{0pc}
\tablecaption{Adaptive Optics Measurements of \hstar \label{tab:ao}}
\tablehead{
\colhead{HJD} &
\colhead{$\rho$} &
\colhead{$\theta$} &
\colhead{$\Delta m$} &
\colhead{}
\\
\colhead{(2,400,000$+$)} &
\colhead{(mas)} &
\colhead{(degree)} &
\colhead{(mag)} &
\colhead{Filter}
}
\startdata
56326.8063 & $55.2 \pm 0.7$    &  $152.72 \pm 0.6$\phn   &  $0.36 \pm 0.09$    &  $J$ \\     
56326.8040 & $55.48 \pm 0.23$  &  $152.2 \pm 0.6$    &  $0.28 \pm 0.06$    &  $H$ \\     
56326.8017 & $55.3 \pm 0.5$    &  $152.6 \pm 0.6$    &  $0.310 \pm 0.023$  &  $K_{\rm S}$ \\
58790.5065 & $55.7 \pm 0.9$    &  $315.5 \pm 1.6$    &  $0.36 \pm 0.09$    &  $J$ \\     
58790.5041 & $54.9 \pm 0.5$    &  $313.9 \pm 1.1$    &  $0.34 \pm 0.05$    &  $H$ \\     
58790.5013 & $54.8 \pm 0.5$    &  $313.6 \pm 0.4$    &  $0.322 \pm 0.020$  &  $K^{\prime}$
\enddata

\end{deluxetable}
\setlength{\tabcolsep}{6pt}  

\begin{figure}
\epsscale{1.05}
\plotone{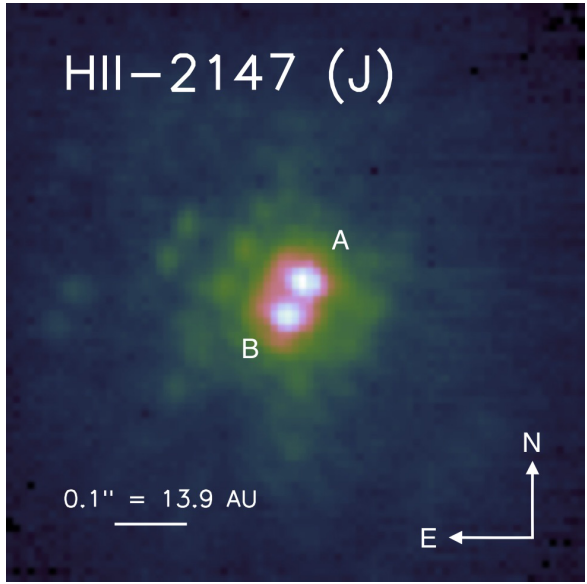}

\figcaption{$J$-band image of \hstar\ from our 2013 NIRC2 adaptive optics
  observations, showing the inner $0\farcs8 \times 0\farcs8$
  region. The fainter star is to the SE at this epoch. \label{fig:ao}}

\end{figure}

We derived final $(x,y)$ coordinates and the magnitude difference of
the two components using the least-squares minimization package {\tt
  MPFIT} in IDL \citep{Markwardt:2009}. We converted NIRC2 pixel
values into sky coordinates using the same methods as described in
\cite{Dupuy:2016} and \cite{Dupuy:2017}, with the only difference
being that we reversed the sign of the P.A.\ offset of $0\fdg252$ in
the \citet{Yelda:2010} calibration, as noted by \cite{Bowler:2018}. We
report the mean of the separation, P.A., and magnitude difference
derived from the set of images taken in each of the three filters (see
Table~\ref{tab:ao}), and we adopt errors based on the rms of
measurements from each data set, with systematic uncertainties in the
astrometric calibration (e.g., $0\fdg009$ in P.A.) added in quadrature
to these rms values for the final errors. An image of \hstar\ in the
$J$ band is shown in Figure~\ref{fig:ao}.

The 2013 observations clearly identify the fainter component as the
one to the SE, whereas by 2019 orbital motion had moved the fainter
star to the NW. The angular separations happen to be similar at both
epochs.  The measured magnitude differences are consistent between the
two sets of observations, in all three near-infrared bandpasses.

If both components of \hstar\ are single main-sequence stars, the less
massive one should be fainter. On this basis, we tentatively identify
the star whose velocities we measured (which we indicated earlier is
less massive than its companion) as the one to the SE in 2013 (NW in
2019), and we refer to it hereafter as `star B', or the `secondary'.
The other component, whose lines we do not see, will be `star A'.

\section{Orbital analysis}
\label{sec:analysis}

The radial-velocity measurements from all sources were combined with
the astrometry from VLBI and our AO imaging into a global analysis to
determine the orbital elements of \hstar.  The parameters we solved
for are the following: the orbital period ($P$), the angular semimajor
axis of the relative orbit ($a\arcsec$), the eccentricity parameters
$\sqrt{e}\cos\omega_{\rm A}$ and $\sqrt{e}\sin\omega_{\rm A}$ (where
$e$ is the eccentricity and $\omega_{\rm A}$ the longitude of
periastron of the primary), the cosine of the orbital inclination
angle ($\cos i$), the position angle of the ascending node for equinox
J2000 ($\Omega_{\rm J2000}$), a reference time of periastron passage
($T$), the center-of-mass velocity of the system ($\gamma$), and the
velocity semiamplitude of the spectroscopically visible secondary star
($K_{\rm B}$). Inclusion of the VLBI data adds the following free
parameters: the parallax ($\pi$), the barycenter proper motion
components ($\mu_{\alpha}^* \equiv \mu_{\alpha} \cos\delta$ and
$\mu_{\delta}$), and offsets ($\Delta\alpha^* \equiv \Delta\alpha
\cos\delta$ and $\Delta\delta$) between the position of the barycenter
at the average time of the VLBI observations ($t_0 =
2,\!456,\!257.119$ HJD = 2012.9011) and a reference position taken to
be the average of the measured VLBI positions ($\alpha_0 =
57\fdg2755330891$, $\delta_0 = +23\fdg7810742666$). Parallax factors
were calculated using the position of the Earth's center as provided
by the JPL Horizons web interface.\footnote{
  \url{https://ssd.jpl.nasa.gov/horizons.cgi}} In addition, we solved
for a possible zero-point offset between the CORAVEL and CfA
velocities ($\Delta RV_{\rm COR}$), and another between the CfA and
Lick velocities ($\Delta RV_{\rm Lick}$). Both of these offsets are to
be added to the velocities from the corresponding data sets in order
to place them on the CfA system.

A further constraint on the orbital elements is provided, in
principle, by the proper motion measured by {\it Gaia}. This will in
general be different from the proper motion of the center of mass,
because of the acceleration caused by the stars moving around each
other \citep[see, e.g.,][]{Brandt:2018}. However, our attempts to
incorporate this constraint resulted in significant tension with the
VLBI measurements, and a poorer fit. Consequently, we have elected not
to use this measurement.\footnote{We note that a comment associated
  with this {\it Gaia}/DR2 entry in the VizieR catalog
  \citep{Ochsenbein:2000} indicates there was a duplicate source in
  the original data reduction that was discarded. This may indicate
  observational, cross-matching, or processing problems, possibly
  compromising the astrometric or photometric results. We believe this
  may explain the difficulty we pointed out.}

Our method of analysis used the {\tt emcee\/}\footnote{\url
  http://dfm.io/emcee/current} code of \cite{Foreman-Mackey:2013},
which is a Python implementation of the affine-invariant MCMC ensemble
sampler proposed by \cite{Goodman:2010}. We used 100 walkers with
chain lengths of 20,000 each, after discarding the burn-in. Uniform
(non-informative) priors over suitable ranges were adopted for all of
the above parameters, and convergence of the chains was checked
visually, requiring also a Gelman-Rubin statistic of 1.05 or smaller
for each parameter \citep{Gelman:1992}.

The relative weighting between the different data sets (CORAVEL, CfA,
Lick, VLBI) was handled by including additional adjustable parameters
to rescale the observational errors as needed to achieve reduced
$\chi^2$ values near unity. For the velocities those parameters were
taken to be multiplicative factors $f_{\rm COR}$, $f_{\rm CfA}$, and
$f_{\rm Lick}$ with uniform priors; for the VLBI measurements, whose
internal errors appeared from a preliminary analysis to be
underestimated, they were ``jitter'' terms $\sigma_{\alpha}$ and
$\sigma_{\delta}$ to be added quadratically to the internal errors,
with log-uniform priors.\footnote{The measured VLBI positions are
  affected by additional systematic errors stemming from the
  uncertainty in the position of the phase-referencing calibrator (the
  quasar J0347+2339; see Section~\ref{sec:vlbi}). However, as we are
  only interested in changes in position, those systematic errors have
  no impact on our analysis.}  All of these terms were solved for
self-consistently and simultaneously with the other orbital quantities
\citep[see][]{Gregory:2005}.  Precession corrections to J2000.0 for
the position angles of the AO measurements are very small, but were
applied nevertheless for completeness.

The results of the MCMC analysis are given in the second column of
Table~\ref{tab:mcmc} (Solution 1), where we report the mode of the
posterior distribution for each parameter along with the 68.3\%
credible intervals. The orbital period corresponds to $18.18 \pm 0.11$
years. The bottom of the table presents derived properties including
the total mass of the system, $M_{\rm tot}$, the masses of the two
components ($M_{\rm A}$ and $M_{\rm B}$) and their mass ratio ($q
\equiv M_{\rm B}/M_{\rm A}$), the distance, the linear semimajor axis
of the orbit ($a$), and the inferred velocity semi-amplitude of the
unseen primary component ($K_{\rm A}$). These derived quantities were
computed by directly combining the chains of the adjusted quantities
involved.

\setlength{\tabcolsep}{5pt}
\begin{deluxetable}{lcc}
\tablewidth{0pc}
\tablecaption{Results from our Combined MCMC Analysis for \hstar \label{tab:mcmc}}
\tablehead{
\colhead{~~~~~~~~~~~~Parameter~~~~~~~~~~~~} &
\colhead{Solution 1} &
\colhead{Solution 2}
\\
\colhead{} & 
\colhead{} &
\colhead{(Adopted)}
}
\startdata
 $P$ (days)\dotfill                     &  $6641^{+40}_{-37}$                 & $6641^{+42}_{-39}$ \\ [1ex]
 $a\arcsec$ (mas)\dotfill               &  $62.33^{+0.45}_{-0.42}$            & $62.32^{+0.45}_{-0.41}$ \\ [1ex]
 $\sqrt{e}\cos\omega_{\rm A}$\dotfill   &  $-0.074^{+0.023}_{-0.022}$         & $-0.074^{+0.023}_{-0.023}$ \\ [1ex]
 $\sqrt{e}\sin\omega_{\rm A}$\dotfill   &  $-0.316^{+0.020}_{-0.018}$         & $-0.316^{+0.021}_{-0.018}$ \\ [1ex]
 $\cos i$\dotfill                       &  $0.2454^{+0.0076}_{-0.0077}$       & $0.2460^{+0.0075}_{-0.0075}$ \\ [1ex]
 $\Omega_{\rm J2000}$ (degree)\dotfill  &  $141.81^{+0.20}_{-0.18}$           & $141.80^{+0.19}_{-0.18}$ \\ [1ex]
 $T$ (HJD$-$2,400,000)\dotfill          &  $47288^{+88}_{-105}$               & $47284^{+93}_{-107}$ \\ [1ex]
 $\gamma$ (\kms)\dotfill                &  $+5.67^{+0.18}_{-0.18}$            & $+5.70^{+0.17}_{-0.17}$ \\ [1ex]
 $K_{\rm B}$ (\kms)\dotfill             &  $7.33^{+0.26}_{-0.26}$             & $7.102^{+0.081}_{-0.081}$ \\ [1ex]
 $\pi$ (mas)\dotfill                    &  $7.312^{+0.027}_{-0.027}$          & $7.310^{+0.026}_{-0.026}$ \\ [1ex]
 $\mu_{\alpha}^*$ (mas yr$^{-1}$)\dotfill & $+19.01^{+0.17}_{-0.17}$          & $+18.855^{+0.039}_{-0.039}$ \\ [1ex]
 $\mu_{\delta}$ (mas yr$^{-1}$)\dotfill &  $-44.74^{+0.14}_{-0.14}$           & $-44.66^{+0.11}_{-0.10}$ \\ [1ex]
 $\Delta\alpha^*$ (mas)\dotfill         &  $-2.44^{+0.50}_{-0.50}$            & $-1.972^{+0.035}_{-0.035}$ \\ [1ex]
 $\Delta\delta$ (mas)\dotfill           &  $+4.09^{+0.89}_{-0.89}$            & $+3.256^{+0.074}_{-0.074}$ \\ [1ex]
 $\Delta RV_{\rm COR}$ (\kms)\dotfill   &  $+0.76^{+0.34}_{-0.34}$            & $+0.73^{+0.35}_{-0.35}$ \\ [1ex]
 $\Delta RV_{\rm Lick}$ (\kms)\dotfill  &  $+0.11^{+0.40}_{-0.40}$            & $+0.04^{+0.40}_{-0.37}$ \\ [1ex]
 $f_{\rm CfA}$ (\kms)\dotfill           &  $1.23^{+0.16}_{-0.12}$             & $1.21^{+0.16}_{-0.12}$ \\ [1ex]
 $f_{\rm COR}$ (\kms)\dotfill           &  $1.23^{+0.16}_{-0.12}$             & $1.29^{+0.15}_{-0.12}$ \\ [1ex]
 $f_{\rm Lick}$ (\kms)\dotfill          &  $1.88^{+0.77}_{-0.36}$             & $1.87^{+0.74}_{-0.39}$ \\ [1ex]
 $\sigma_{\alpha}$ (mas)\dotfill        &  $0.075^{+0.022}_{-0.015}$          & $0.071^{+0.021}_{-0.014}$ \\ [1ex]
 $\sigma_{\delta}$ (mas)\dotfill        &  $0.201^{+0.059}_{-0.039}$          & $0.195^{+0.060}_{-0.036}$ \\ [1ex]
\hline \\ [-1.5ex]
\multicolumn{3}{c}{Derived quantities} \\ [1ex]
\hline \\ [-1.5ex]
 $e$\dotfill                            &  $0.106^{+0.011}_{-0.010}$          & $0.105^{+0.011}_{-0.011}$   \\ [1ex]
 $\omega_{\rm A}$ (degree)\dotfill      &  $257.0^{+4.1}_{-4.5}$              & $256.8^{+4.2}_{-4.6}$        \\ [1ex]
 $i$ (degree)\dotfill                   &  $75.79^{+0.45}_{-0.45}$            & $75.78^{+0.43}_{-0.46}$     \\ [1ex]
 Distance (pc)\dotfill                  &  $136.75^{+0.50}_{-0.50}$           & $136.78^{+0.50}_{-0.46}$     \\ [1ex]
 $a$ (au)\dotfill                       &  $8.523^{+0.071}_{-0.066}$          & $8.525^{+0.070}_{-0.064}$ \\ [1ex]
 $K_{\rm A}$ (\kms)\dotfill             &  $6.28^{+0.25}_{-0.26}$             & $6.511^{+0.075}_{-0.074}$   \\ [1ex]
 $M_{\rm tot}$ ($M_{\sun}$)\dotfill     &  $1.873^{+0.046}_{-0.043}$          & $1.875^{+0.045}_{-0.045}$ \\ [1ex]
 $M_{\rm A}$ ($M_{\sun}$)\dotfill       &  $1.008^{+0.043}_{-0.040}$          & $0.978^{+0.024}_{-0.024}$ \\ [1ex]
 $M_{\rm B}$ ($M_{\sun}$)\dotfill       &  $0.865^{+0.040}_{-0.040}$          & $0.897^{+0.022}_{-0.022}$ \\ [1ex]
 $q \equiv M_{\rm B}/M_{\rm A}$\dotfill &  $0.856^{+0.065}_{-0.060}$          & $0.9168^{+0.0039}_{-0.0039}$
\enddata

\tablecomments{Solution~2 incorporates a prior on the mass ratio
  $q_{\rm AB}$, derived from the NIR magnitude differences (see
  text). For both solutions, the parameter values listed correspond to
  the mode of the respective posterior distributions, and the
  uncertainties represent the 68.3\% credible intervals.}

\end{deluxetable}
\setlength{\tabcolsep}{6pt}

Appealing to a 125~Myr solar-metallicity model isochrone for the
Pleiades from the PARSEC series \citep{Chen:2014}, we find that a star
with the mass of the secondary ($M_{\rm B} = 0.865~M_{\sun}$) is
expected to have an effective temperature of about 5220~K, which is
somewhat lower than our spectroscopically measured value ($5390 \pm
100$~K). A further check on the accuracy of our solution can be made
using the measured masses to predict the primary/secondary magnitude
differences in the NIR. These can then be compared against the
measured values from Table~\ref{tab:ao}. Similarly, we can calculate
the expected total system brightness at both optical and NIR
wavelengths, for comparison with the magnitudes in the {\it Gaia}/DR2
and 2MASS catalogs. For these tests we have preferred not to rely
entirely on the model isochrone used above to predict fluxes, as we
find that it does not match the empirical color-magnitude diagram of
the cluster sufficiently well for our purposes. Instead, we developed
semi-empirical relations to predict fluxes in the {\it Gaia\/} and
2MASS bandpasses as a function of mass. Briefly, we fit spline
relations to the empirical color-magnitude diagram of the Pleiades,
and then used the models to provide the connection between masses and
observed bandpass magnitudes in the two photometric systems.  The
details of this derivation are given in the Appendix, where we also
present a test of the accuracy of these relations using the other
binary systems in the Pleiades that have dynamically measured masses.

Using these semi-empirical relations, we find that the expected 2MASS
magnitudes from our Solution~1 for the combined light of \hstar\ agree
quite well with the measured $JHK_{\rm S}$ values (see
Table~\ref{tab:mags}). The corresponding predictions for the {\it
  Gaia} bandpasses are also formally consistent within the
uncertainties, although they are all systematically brighter than the
measured magnitudes by about 0.1~mag, on average. On the other hand,
the predicted $JHK$ magnitude differences between stars A and B are
all larger than observed by 0.2--0.3~mag, However, the uncertainties
in this case are large enough that there is again formal consistency
with the measurements.  The inflated uncertainties for $\Delta J$,
$\Delta H$, and $\Delta K$ are a reflection of the relatively
imprecise value of the mass ratio, which is what largely determines
those magnitude differences.

\setlength{\tabcolsep}{3.5pt}
\begin{deluxetable}{lccc}
\tablewidth{0pc}
\tablecaption{Predicted Brightness Measurements for \hstar\ from our
  MCMC Solutions in Table~\ref{tab:mcmc} \label{tab:mags}}
\tablehead{
\colhead{~~~~Parameter~~~~} &
\colhead{Solution 1} &
\colhead{Solution 2} &
\colhead{Measurements} \\
\colhead{} &
\colhead{} &
\colhead{(Adopted)} &
\colhead{}
}
\startdata
 $G$ (mag)\dotfill           & $10.32 \pm 0.17$\phn   & $10.39 \pm 0.15$\phn   & $10.421 \pm 0.030$\phn \\
 $G_{\rm BP}$ (mag)\dotfill  & $10.72 \pm 0.19$\phn   & $10.81 \pm 0.17$\phn   & $10.873 \pm 0.030$\phn \\
 $G_{\rm RP}$ (mag)\dotfill  & $9.75 \pm 0.14$        & $9.81 \pm 0.13$        & $9.832 \pm 0.030$ \\ [1ex]
 $J$ (mag)\dotfill           & $9.10 \pm 0.10$        & $9.13 \pm 0.10$        & $9.166 \pm 0.021$ \\
 $H$ (mag)\dotfill           & $8.733 \pm 0.086$      & $8.750 \pm 0.083$      & $8.719 \pm 0.042$ \\
 $K_{\rm S}$ (mag)\dotfill   & $8.633 \pm 0.083$      & $8.649 \pm 0.080$      & $8.603 \pm 0.017$ \\ [1ex]
 $\Delta J$ (mag)\dotfill    & $0.68 \pm 0.32$        & $0.380 \pm 0.019$      & $0.360 \pm 0.064$ \\
 $\Delta H$ (mag)\dotfill    & $0.57 \pm 0.27$        & $0.317 \pm 0.016$      & $0.310 \pm 0.042$ \\
 $\Delta K$ (mag)\dotfill    & $0.55 \pm 0.26$        & $0.306 \pm 0.015$      & $0.316 \pm 0.015$
\enddata

\tablecomments{Solution~2 uses a prior on the mass ratio (see text).
  The magnitude differences in the last three rows of the last column
  are averages of the values in Table~\ref{tab:ao} in each bandpass.
  Uncertainties for the {\it Gaia\/} magnitudes have been arbitrarily
  increased, to account for possible biases in the models used to
  derive the semi-empirical relations mentioned in the text (see the
  Appendix).}

\end{deluxetable}
\setlength{\tabcolsep}{6pt}

\begin{figure}
\epsscale{1.15}
\plotone{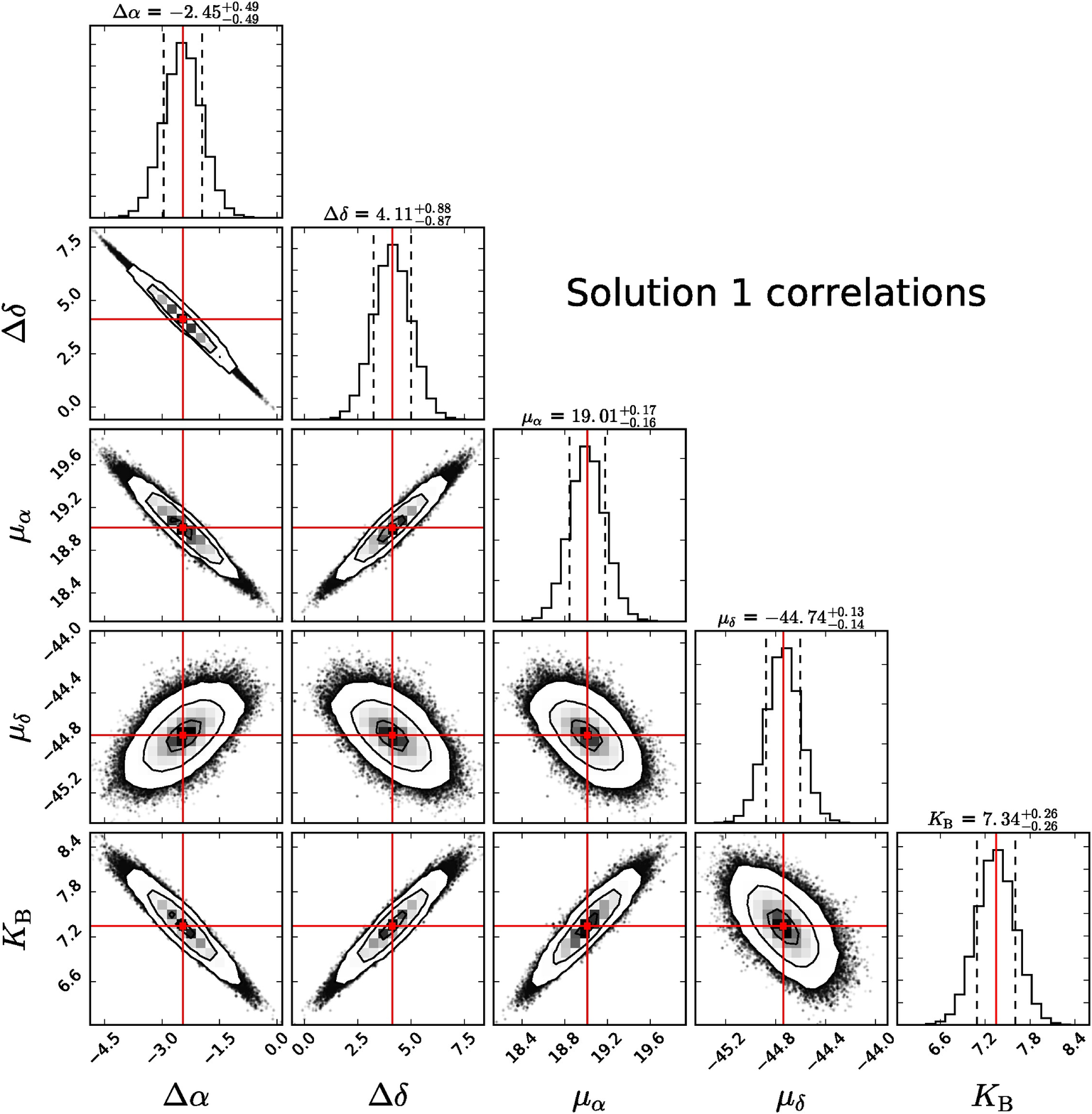}
\vskip 5pt
\plotone{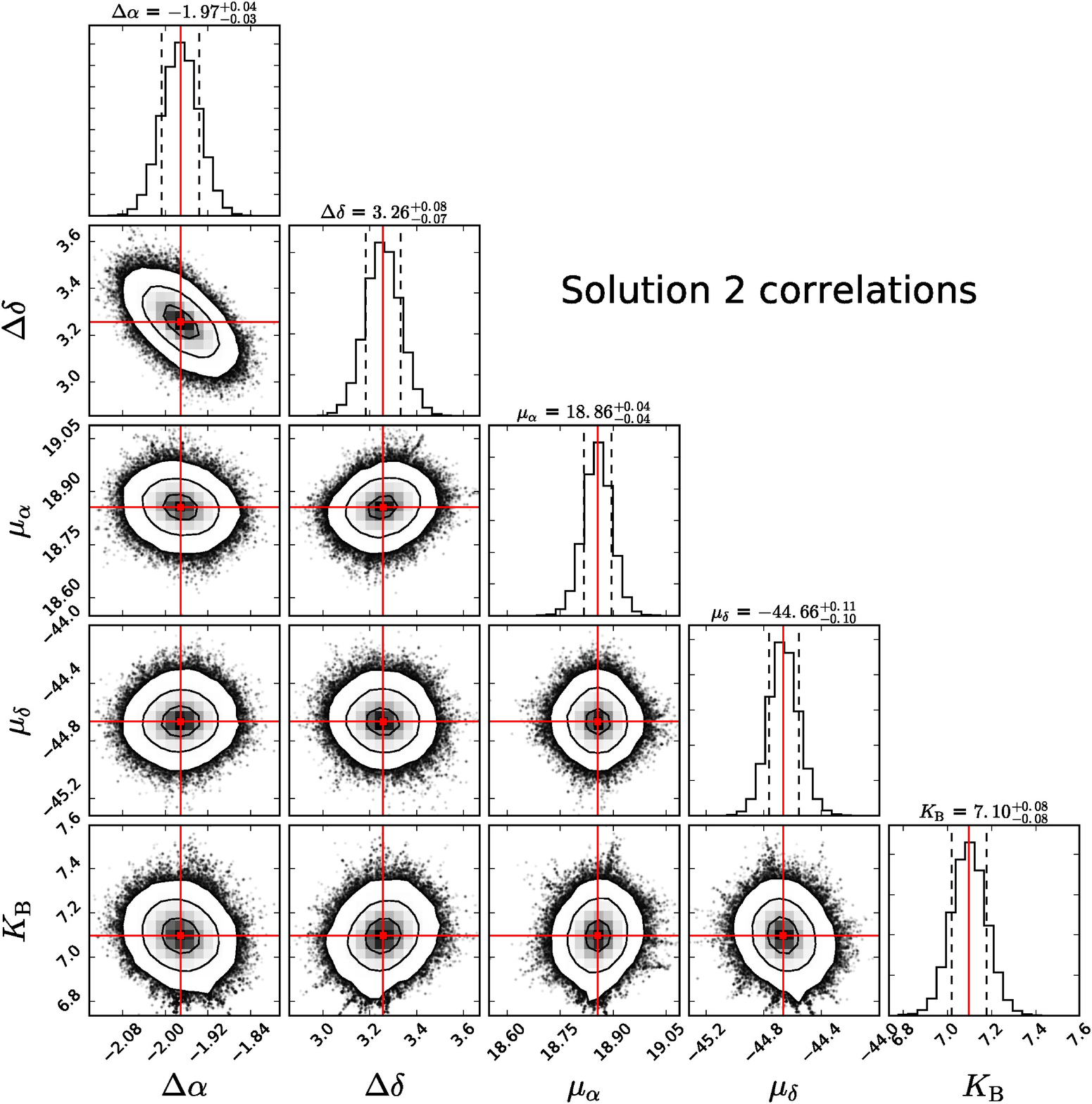}

\figcaption{\emph{Top:} Corner plot
  \citep{Foreman-Mackey:2016}\footnote{\url{
      https://github.com/dfm/corner.py}} showing the correlations
  among the parameters from Solution~1 that constrain the mass ratio
  $q$ the most, and which happen to be the ones most strongly
  correlated amongst themselves. The contours correspond to the 1, 2,
  and 3$\sigma$ confidence levels. \emph{Bottom:} Same as above, for
  Solution~2. Mutual correlations are greatly reduced through the use
  of photometric constraints (see text). \label{fig:corner}}

\end{figure}

The fact that the predicted $JHK$ magnitude differences are worse, in
a systematic sense, than those for the combined light suggests the
possibility of a small bias in the mass ratio $q$.  The only
observations used in our analysis that constrain $q$ are the VLBI
measurements for the primary and secondary, which are measured
separately on an absolute reference frame. The constraint is weak,
however, because the VLBI observations only cover a small fraction of
the orbit. The accuracy of the mass ratio will then depend critically
on how well the center of mass can be located on the plane of the sky,
as represented by the free parameters $\Delta\alpha^*$ and
$\Delta\delta$. These two variables happen to be the ones most
strongly correlated amongst themselves and with other free parameters,
so it would not be surprising if they were affected by subtle biases.
Indeed, $\Delta\alpha^*$ correlates very strongly with $\Delta\delta$
(correlation coefficient $-0.998$), with $\Delta\mu_{\alpha}^*$
($-0.965$), and with $K_{\rm B}$ ($-0.950$), whereas $\Delta\delta$
shows significant correlation with $\Delta\mu_{\alpha}^*$ (+0.965) and
$K_{\rm B}$ ($+0.950$). This is illustrated in the top panel of
Figure~\ref{fig:corner}.

\begin{figure}
\epsscale{1.15}
\plotone{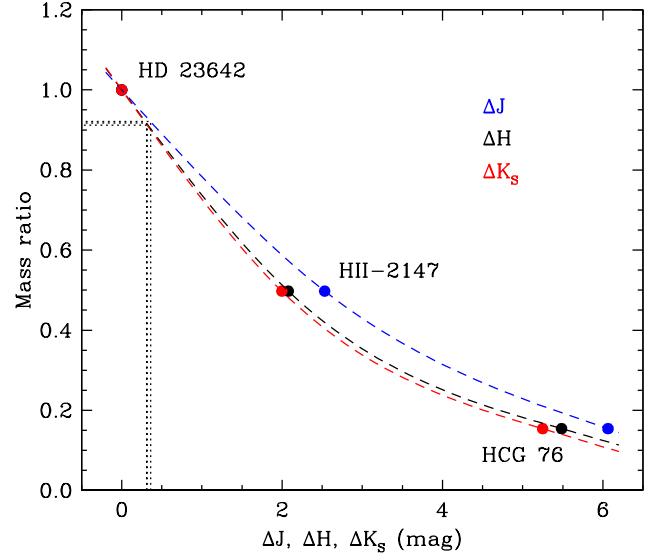}

\figcaption{Empirical diagram of the mass ratio as a function of the
  $JHK$ magnitude differences, based on the total masses and 2MASS
  magnitudes of the Pleiades binaries HD~23642 and HCG~76
  \citep{David:2016}, as well as \hstar\ (total mass from
  Table~\ref{tab:mcmc}, Solution~1).  HD~23642 was used as the
  reference, so its mass ratio is unity and its magnitude differences
  are zero with respect to itself.  Interpolating spline curves are
  shown in each filter, and dotted lines illustrate how we derived
  three estimates of $q$ for \hstar\ from the measured $\Delta J$,
  $\Delta H$, and $\Delta K$ values provided by our AO imaging
  (average for the two epochs in Table~\ref{tab:ao}). The mean mass
  ratio we obtain is $q = 0.917 \pm 0.004$.\label{fig:qprior}}

\end{figure}

The availability of the near-infrared magnitude differences from our
AO observations presents an opportunity to check or improve the
accuracy of the mass ratio, as they are the measurements with the
closest connection to $q$.  To this end, we constructed an empirical
relation that allows us to predict $q$ from a difference in brightness
in the $JHK$ bandpasses. For this, we enlisted the other binaries in
the Pleiades that have measured masses. As none of them have NIR
brightness measurements for the individual components, we opted
instead to use their total system masses along with their
combined-light 2MASS magnitudes.  Only two of the three available
systems have reliable 2MASS magnitudes \citep[HD~23642 and
  HCG~76;][]{David:2016}, with the third (Atlas, $V = 3.6$) being too
bright. To these we therefore added \hstar\ itself, and used its total
mass from Solution~1 together with its combined $JHK_{\rm S}$
magnitudes from the 2MASS catalog. We then calculated the ratio of the
total masses of \hstar\ and HCG~76 with respect to the total mass of
HD~23642 (the more massive system), and the magnitude differences in
$JHK$ between each of the two lighter systems and HD~23642. A diagram
of these three system mass ratios and magnitude differences is shown
in Figure~\ref{fig:qprior}, with corresponding interpolated spline
curves for each filter.  Finally, with the measured values of $\Delta
J$, $\Delta H$, and $\Delta K$ for \hstar\ from Table~\ref{tab:ao}
(averaged between the two AO epochs), we used these curves to infer
three values for the mass ratio. The mean and standard deviation are
$q = 0.917 \pm 0.004$.

We then carried out another orbital solution using the same
observations as before, this time applying the above result as a
Gaussian prior on the mass ratio.  We refer to this as Solution~2. The
orbital elements and derived properties we obtained are listed in the
last column of Table~\ref{tab:mcmc}. The results are all very similar
to those of our initial analysis, with the anticipated exception of
the parameters that were previously highly correlated with each other
(see above). For those, the uncertainties are now considerably
reduced, as are the mutual correlations (see Figure~\ref{fig:corner},
bottom panel).  The mass ratio is slightly larger in the new solution,
but more than an order of magnitude more precise.  As a result, the
individual masses are also somewhat different and considerably better
determined, whereas the total mass is essentially unchanged compared
to its uncertainty. The predicted combined magnitudes in the {\it
  Gaia\/} bandpasses agree better than before with the observations,
as can be seen in Table~\ref{tab:mags}, while the magnitude
differences now track the measured values closely, by construction.

The mass of star~B in this new solution is now slightly larger than
before, and the corresponding effective temperature according to the
PARSEC isochrone is 5350~K, which is much closer to the value we
measured spectroscopically. We conclude that this set of orbital
parameters is consistent with all available observational constraints
for \hstar.

Our solution leads to a parallax for \hstar\ of $\pi = 7.310 \pm
0.026$~mas, corresponding to a linear distance of
$136.78^{+0.50}_{-0.46}$~pc. This is twice as precise as, but in good
agreement with the parallax reported in the {\it Gaia}/DR2 catalog,
after that value is adjusted for a systematic difference compared to
VLBI determinations ($+0.075 \pm 0.029$~mas) following \cite{Xu:2019},
giving $\pi_{\rm Gaia} = 7.284 \pm 0.059$~mas.

\begin{figure}
\epsscale{1.15}
\plotone{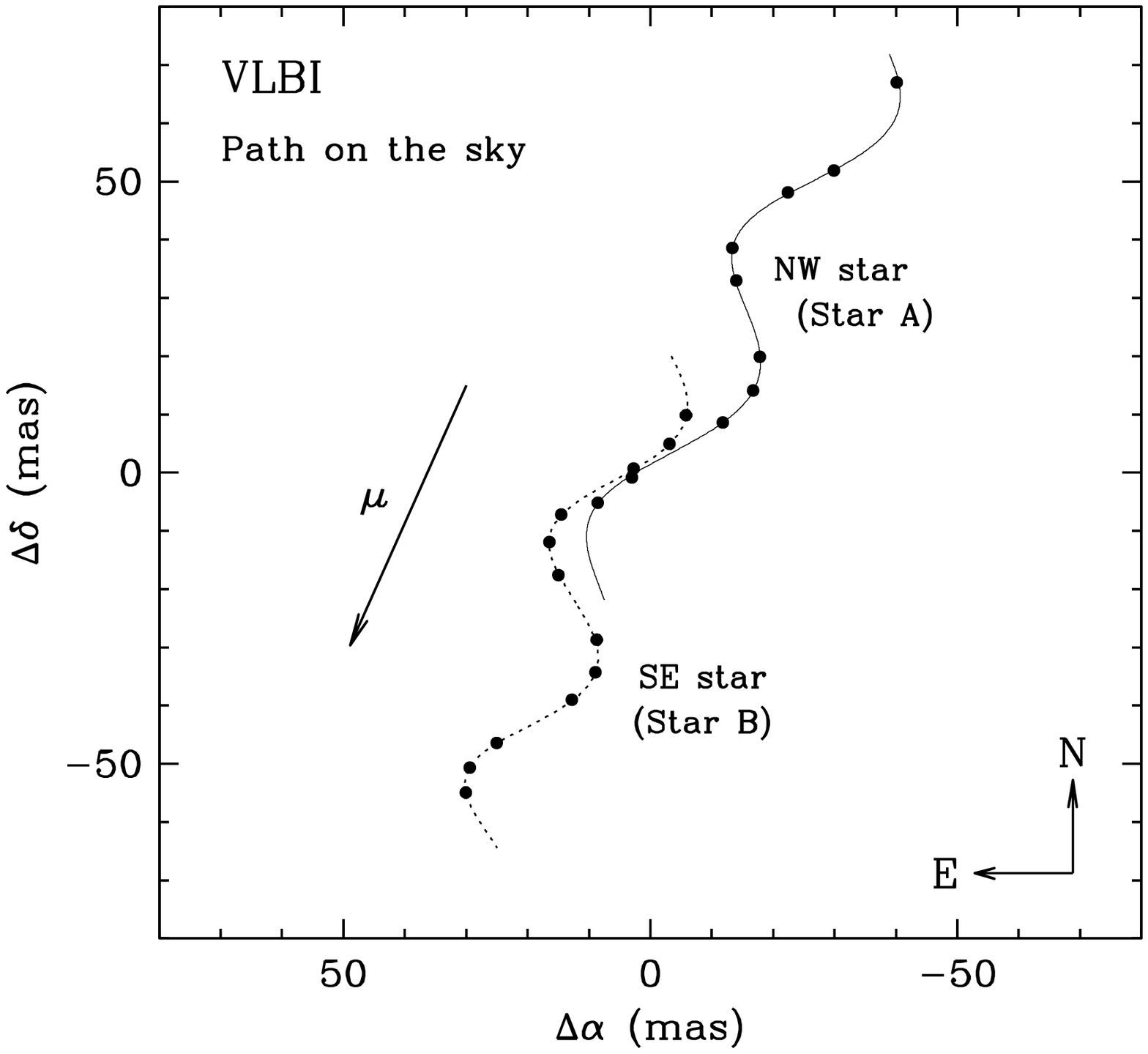}
\vskip 5pt
\plotone{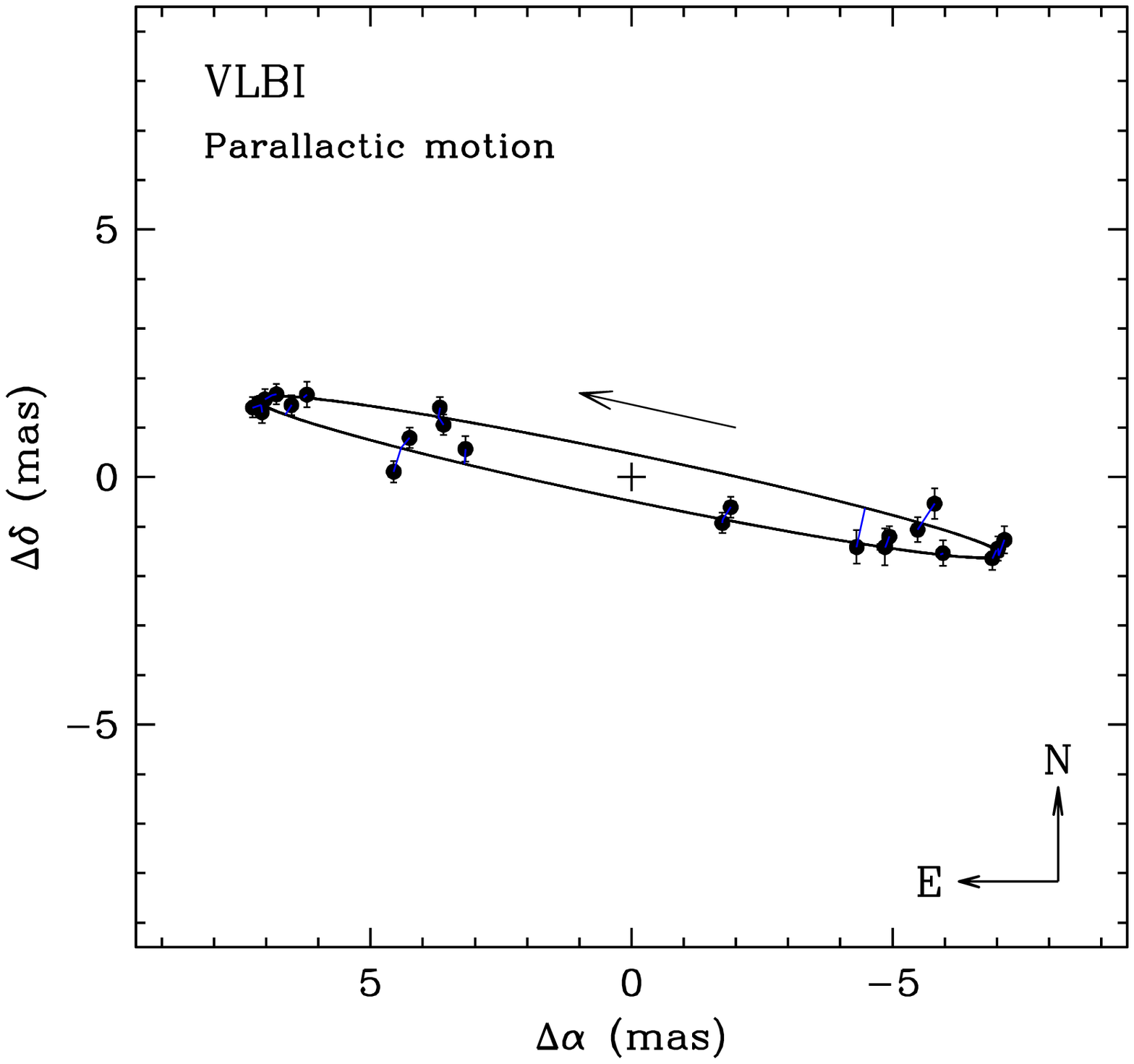}

\figcaption{\emph{Top:} Trajectory of each component of \hstar\ on the
  plane of the sky from our Solution~2 of Table~\ref{tab:mcmc}, along
  with the VLBI observations (error bars are smaller than the points).
  The origin of the coordinate system is taken to be the average of
  all VLBI positions ($\alpha_0 = 57\fdg2755330891$, $\delta_0 =
  +23\fdg7810742666$), and the arrow shows the change in position in
  one year due to proper motion. \emph{Bottom:} Parallactic motion,
  after removal of the orbital motion and proper motion from the VLBI
  measurements. The error bars in right ascension are smaller than the
  symbol size. Observations are connected with the predicted positions
  by a thin line. \label{fig:path}}

\end{figure}

\begin{figure}
\epsscale{1.15}
\plotone{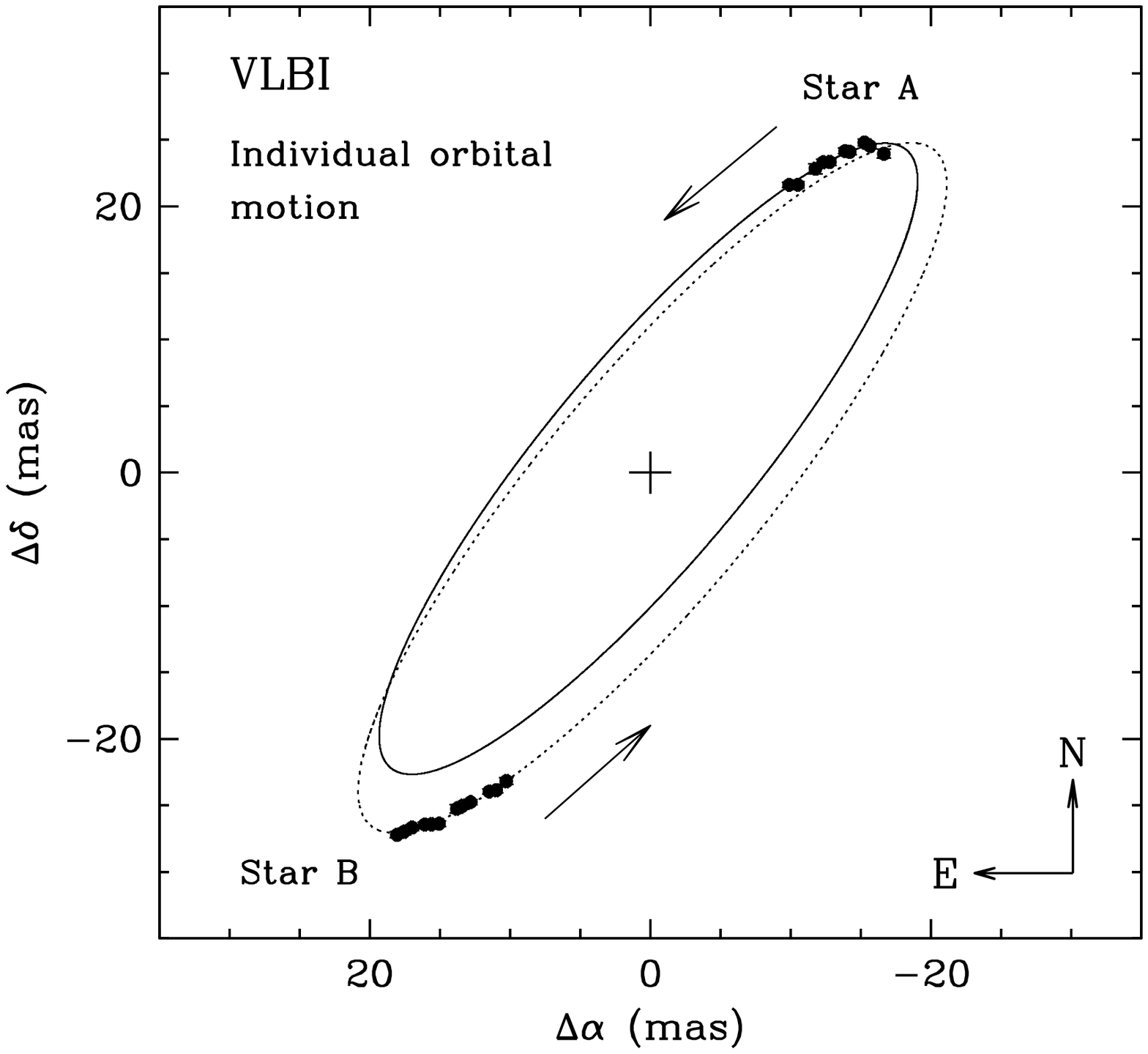}
\vskip 5pt
\plotone{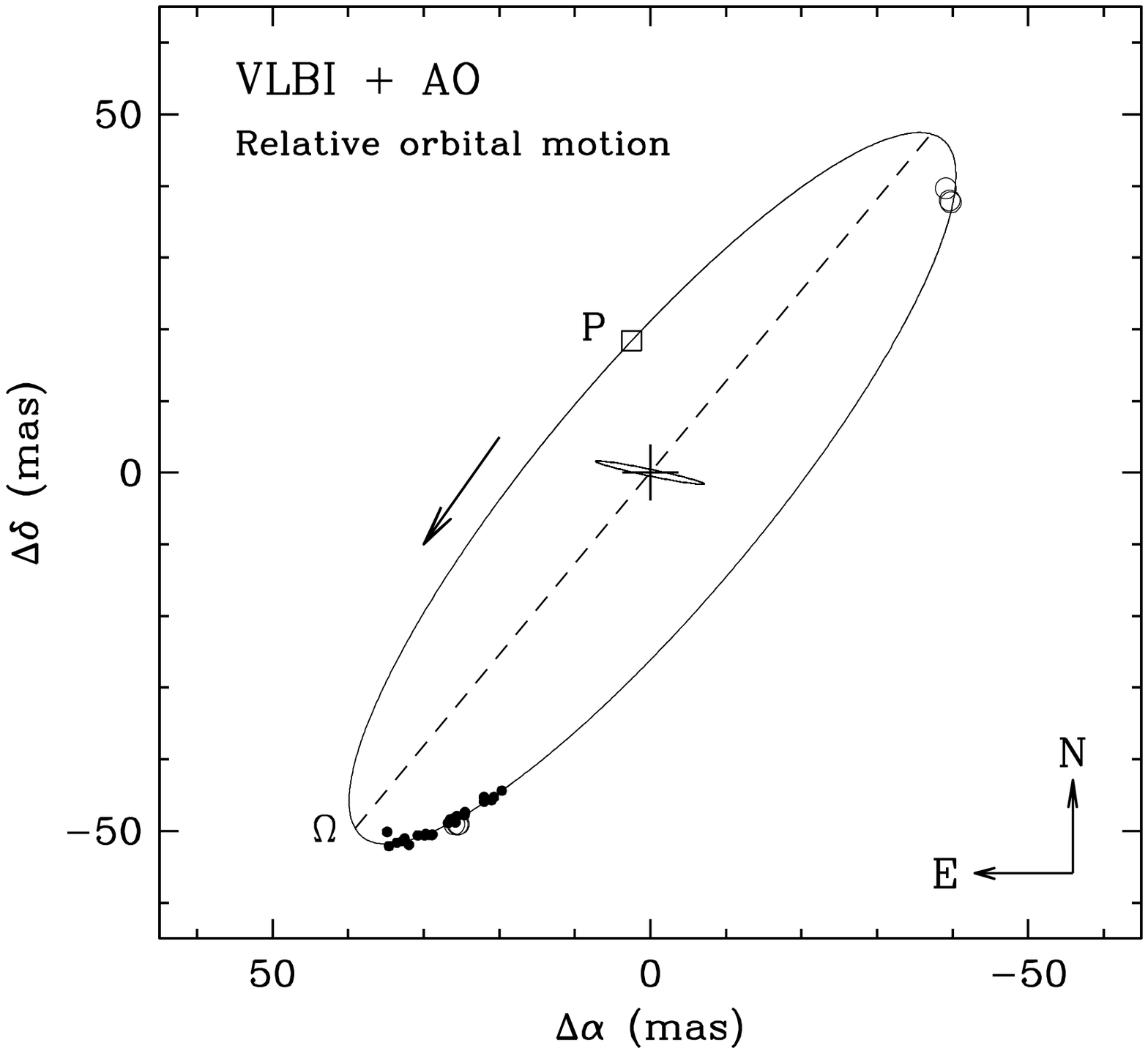}

\figcaption{\emph{Top:} VLBI measurements for each component of
  \hstar\ in their orbit around the center of mass, along with our
  models from Solution~2 in Table~\ref{tab:mcmc}. Proper motion and
  parallactic motion have been removed. \emph{Bottom:} Relative orbit
  of \hstar\ from our Solution~2 model (star~A at the center), along
  with the VLBI observations projected onto the orbit (filled
  circles). Error bars are smaller than the symbols. Our adaptive
  optics observations from 2013 and 2019 are shown with open circles.
  The dashed line marks the line of nodes (ascending node at the
  bottom), and the small ellipse at the center represents the
  parallactic motion to scale, from Figure~\ref{fig:path}. Periastron
  is marked with a ``P'', and motion is direct
  (arrow).\label{fig:vlbiorbit}}

\end{figure}

\begin{figure}
\epsscale{1.15}
\plotone{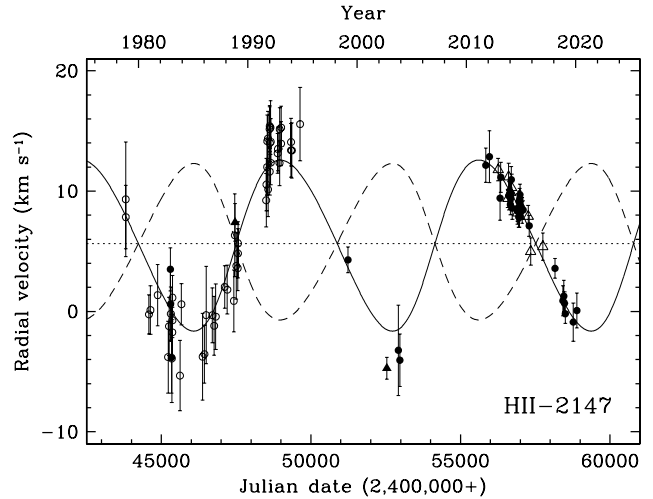}

\figcaption{Radial velocity model from our analysis together with the
  observations of \hstar. Symbols as in Figure~\ref{fig:allrvs}. The
  dashed line represents the predicted velocity curve of the unseen
  spectroscopic secondary, and the dotted line marks the
  center-of-mass velocity of the system.\label{fig:rvorbit}}

\end{figure}

A representation of the path of each star on the plane of the sky
together with the VLBI measurements is shown in the top panel of
Figure~\ref{fig:path}, with the arrow indicating the direction and
magnitude of the change due to proper motion over the span of one
year. The bottom panel shows the parallactic ellipse for \hstar\ along
with the VLBI measurements corrected for proper motion and orbital
motion, following our Solution~2.  The measured position of each
component relative to the barycenter is illustrated in the top panel
of Figure~\ref{fig:vlbiorbit}, with the proper motion and parallactic
motion removed. Motion in the relative orbit (star~B relative to
star~A) is shown in the bottom panel, including the AO measurements.
Orbital motion on the plane of the sky is counterclockwise (direct).
The predicted radial-velocity curves for the primary and secondary can
be seen in Figure~\ref{fig:rvorbit} along with the observations.

\begin{figure}
\epsscale{1.15}
\plotone{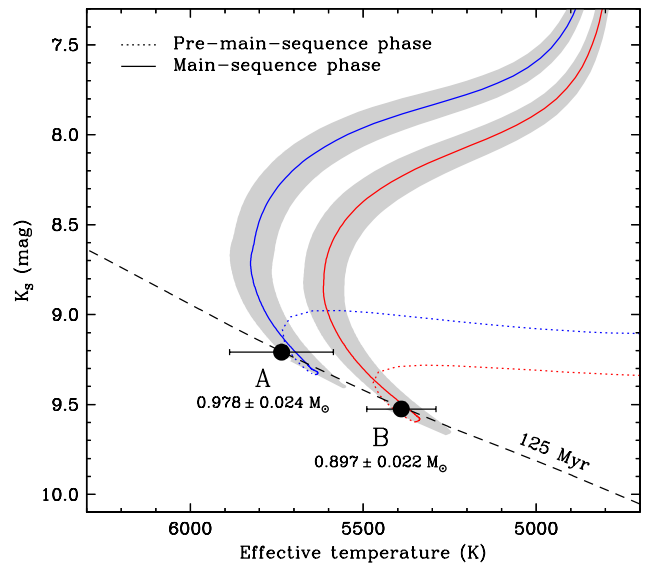}

\figcaption{MIST evolutionary tracks \citep{Choi:2016} for the
  measured masses of \hstar, compared with the observations. Solid
  lines correspond to the main-sequence phase, and dotted ones to the
  pre-main-sequence phase. Shaded areas indicate the uncertainty in
  the location of the tracks that comes from the mass errors. The
  dashed line is a solar-metallicity 125~Myr isochrone for the
  Pleiades, from the same series of models. Model magnitudes have been
  corrected for the distance modulus using the measured parallax of
  \hstar, and for extinction (see Appendix). Individual $K_{\rm S}$
  magnitudes were calculated using the combined $K_{\rm S}$ brightness
  from 2MASS together with $\Delta K$ from our AO measurements
  (Section~\ref{sec:ao}). The temperature for the secondary is
  spectroscopic (Section~\ref{sec:newspectroscopy}); the one for the
  primary was derived from the secondary value and a temperature
  difference based on the AO magnitude differences and the models (see
  text).\label{fig:mist}}

\end{figure}

As a check on our mass determinations for \hstar, we compared the
measurements with stellar evolution models from the MIST series
\citep[MESA Isochrones and Stellar Tracks;][]{Choi:2016}. Most models
including these have difficulty matching the color-magnitude diagram
of the Pleiades, as mentioned earlier, so here we have compared theory
and observations via evolutionary tracks in a diagram of apparent
$K_{\rm S}$-band magnitude versus effective temperature. An estimate
of $T_{\rm eff}$ for the primary was derived from the spectroscopic
value for the secondary and a temperature offset $\Delta T_{\rm eff}$
calculated using the $JHK$ magnitude differences from AO. The offset
was determined by using a 125~Myr MIST isochrone for the Pleiades, and
recording the changes in temperature corresponding to changes in
$JHK_{\rm S}$ brightness from the predicted values for the secondary,
at its measured mass. Because the models are used here only in a
differential sense, the dependence of the temperature offset (and
therefore of the primary temperature) on theory is weak.  The average
of the three $\Delta T_{\rm eff}$ estimates is 350~K, resulting in a
final primary temperature of $5740 \pm 150$~K (conservative
uncertainty). Figure~\ref{fig:mist} indicates that the models are
consistent with the component temperatures and brightnesses at their
measured masses.

\subsection{Alternate scenarios}
\label{sec:alternate}

While Solution~2 provides satisfactory agreement with all astrometric,
spectroscopic, and photometric observations for \hstar, it still
requires an explanation for the lack of detection of the lines of
star~A in our spectra, especially given that it is the more massive
component, and should therefore be brighter. A reasonable explanation
would be rapid rotation, as mentioned earlier.  Another possibility is
that star~A is itself a close binary, composed of stars Aa and Ab.
This could alleviate the mystery somewhat by making the detection of
its spectral lines more challenging, more so if one or both of its
components are also rapid rotators. Under this scenario, component~A
can actually be fainter than component~B (the spectroscopically
visible star), contrary to what we have assumed so far. This is
because dividing up its mass among two smaller stars (Aa and Ab) can
reduce its total brightness, depending on the mass ratio.  This could
have significant consequences for the orbital solution, because it
would reverse the location of the star that has the measured
velocities relative to the VLBI measurements, potentially changing
some of the orbital elements. On the other hand, other mass ratios
between Aa and Ab would allow it to remain the brighter component.

In principle, we can explore both of these possible triple-star
scenarios (Aa\,+\,Ab brighter than star B, or vice versa) by adding
the mass ratio $q_{\rm A} \equiv M_{\rm Ab}/M_{\rm Aa}$ as a free
parameter. However, the astrometric, spectroscopic, and photometric
observations used up to now do not constrain this new parameter, so
for this we chose to make use of the {\it Gaia\/} and 2MASS magnitudes
of \hstar\ as measurements, with their corresponding uncertainties. We
used the semi-empirical mass-magnitude relations developed in the
previous section to predict the individual magnitudes at each step of
the iterations, adding the appropriate term to the likelihood
function. As mentioned in the Appendix, we do not expect this mapping
between theoretical masses and fluxes to be free from systematic
error: there are many physical ingredients in the models that can
affect the passband-specific flux predictions in ways that are
difficult to quantify. To guard against this, we took two precautions
in our new solutions: we allowed for a shift $\Delta M$ in the mass
scale of the models by adding it as one more free parameter in our
MCMC analysis (solved simultaneously with the rest), and we
conservatively increased the photometric uncertainties from {\it
  Gaia\/} and 2MASS by adding 0.02~mag in quadrature to the formal
errors, as a way of accounting for biases in the model fluxes as well
as for variable extinction within the Pleiades cluster. The resulting
masses will therefore be model-dependent, to some extent, as opposed
to the ones in Solution~2, which are purely empirical.

We produced two new solutions using the magnitudes and the $JHK$
magnitude differences from Table~\ref{tab:ao} as observables, along
with the same astrometric and spectroscopic information used in
Solution~2 above.  We refer to the new triple-star solution with
component~A being brighter than star~B as Solution~3, and to the one
with component~A being fainter as Solution~4.  Most of the orbital
elements are largely unchanged. We report the results for the masses
in Table~\ref{tab:triple}, with the values from Solution~2 repeated
for reference in the second column.

The total mass of the system is very nearly the same in all three
cases, indicating it is robust no matter what the configuration is.
This is because it is essentially constrained by the orbital period,
the semimajor axis, and the parallax, each of which is well determined
to better than 1\%.  We find that in order to fit the observations,
Solution~3 requires a shift in the scale of the model masses of
$\Delta M = -0.062~M_{\sun}$, which seems uncomfortably large: it
amounts to almost $\sfrac{1}{3}$ of the mass of star~Ab.  Furthermore,
for a star with the mass of $M_{\rm B}$ in this solution, the PARSEC
isochrone for the Pleiades predicts an effective temperature of
4970~K, which is more than 400~K cooler than the value we measured
spectroscopically ($5390 \pm 100$~K). For these reasons we do not
consider this model to be plausible.  Solution~4, in which component~A
is fainter than star~B, fares somewhat better regarding the
temperature, although the predicted value of 5140~K is still 250~K
cooler than we measure. Moreover, the offset required in the mass
scale of the models is even larger than before, $\Delta M =
-0.113~M_{\sun}$, which is again about $\sfrac{1}{3}$ of the mass of
star~Ab in this configuration.

Aside from being more contrived, we conclude that neither of the
triple-star scenarios provides a description of the system as
satisfactory as Solution~2, given the available observational
constraints. Nor do they help in explaining the lack of detection of
the lines of another star in our spectra, given that in both cases
star~Aa would not be too different in mass (and therefore brightness)
from star~B (see Table~\ref{tab:triple}). The two-star scenario
(Solution~2) is thus favored by all available evidence.

\setlength{\tabcolsep}{4pt}
\begin{deluxetable}{lccc}
\tablewidth{0pc}

\tablecaption{Test Solutions of Triple-star Scenarios for
  \hstar, Compared with our Adopted Results from
  Solution~2 \label{tab:triple}}

\tablehead{
\colhead{~~~~~Parameter~~~~} &
\colhead{Solution 2} &
\colhead{Solution 3} &
\colhead{Solution 4}
\\
\colhead{} &
\colhead{(Adopted)} &
\colhead{(A is brighter)} &
\colhead{(B is brighter)}
}
\startdata
 $M_{\rm tot}$ ($M_{\sun}$)\dotfill               & $1.875^{+0.045}_{-0.045}$  & $1.860^{+0.043}_{-0.039}$  & $1.875^{+0.042}_{-0.039}$     \\ [1ex]
 $M_{\rm A}$ ($M_{\sun}$)\dotfill                 & $0.978^{+0.024}_{-0.024}$  & $1.049^{+0.035}_{-0.042}$  & $1.028^{+0.030}_{-0.030}$     \\ [1ex]
 ~~~$M_{\rm Aa}$ ($M_{\sun}$)\dotfill                & \nodata                    & $0.876^{+0.036}_{-0.028}$  & $0.717^{+0.035}_{-0.036}$     \\ [1ex]
 ~~~$M_{\rm Ab}$ ($M_{\sun}$)\dotfill                & \nodata                    & $0.173^{+0.047}_{-0.050}$  & $0.313^{+0.048}_{-0.049}$     \\ [1ex]
 $M_{\rm B}$ ($M_{\sun}$)\dotfill                 & $0.897^{+0.022}_{-0.022}$  & $0.806^{+0.040}_{-0.021}$  & $0.846^{+0.024}_{-0.021}$     \\ [1ex]
 $q_{\rm A}$\dotfill                              & \nodata                    & $0.197^{+0.061}_{-0.057}$  & $0.434^{+0.091}_{-0.080}$     \\ [1ex]
$\Delta M$ ($M_{\sun}$)\dotfill                  & \nodata                    & $-0.062^{+0.036}_{-0.025}$ & $-0.113^{+0.027}_{-0.023}$    \\ [1ex]
 $T_{\rm eff}^{\rm B}$ (K)\dotfill                & 5350                       & 4970                       & 5140                       
\enddata

\tablecomments{Solution~2 is our reference solution from the last
  column of Table~\ref{tab:mcmc}. Solution~3 has the combined light of
  component~A (composed of stars Aa and Ab) being brighter than
  star~B, and in Solution~4 it is the opposite.  $\Delta M$ represents
  a shift applied to the mass scale from the models (see text).
  $T_{\rm eff}^{\rm B}$ is the predicted effective temperature from
  the PARSEC model for the Pleiades for a star with mass $M_{\rm B}$
  (i.e., the spectroscopically visible star). The mass ratio within
  star~A is defined as $q_{\rm A} \equiv M_{\rm Ab}/M_{\rm Aa}$. The
  values listed correspond to the mode of the respective posterior
  distributions, and the uncertainties represent the 68.3\% credible
  intervals.}

\end{deluxetable}
\setlength{\tabcolsep}{6pt}

\section{Rotation}
\label{sec:rotation}

Photometric monitoring of \hstar\ by a number of authors has yielded
several different estimates of the rotation period based on the
modulation due to spots. All are very short, placing the object in the
category of the UFRs in the Pleiades \citep{Hartman:2010,
  Rebull:2016}. \cite{Norton:2007} analyzed SuperWASP photometry, and
reported $P_{\rm rot} = 0.3082$ days with a peak-to-peak amplitude of
about 0.04 mag in unfiltered white light. \cite{Hartman:2010} used
observations from the HATNet transiting planet survey, and gave a
period of $P_{\rm rot} = 0.7762$ days that is 2.5 times longer, with a
total amplitude of 0.031 mag in the Sloan $r$ band. Subsequently,
\cite{Kiraga:2012} measured a period $P_{\rm rot} = 0.3083$ days that
is essentially identical to that of \cite{Norton:2007}, with an
amplitude of about 0.07 mag from the $V$-band ASAS photometry. More
recently, \cite{Rebull:2016} used observations from NASA's \emph{K2\/}
mission, and reported a preferred rotation period of 0.7768 days with
an amplitude of 0.037 mag, as well as a secondary modulation with a
much shorter period of 0.1541 days that is exactly half of the value
found by \cite{Norton:2007} and \cite{Kiraga:2012}. The various
estimates therefore appear to be in the ratios 1:2:5.

If the true spin rate is the fastest one (0.1541 days), it would imply
an equatorial rotational velocity of about 250~\kms\ for a star such
as component B in \hstar. Given our spectroscopically measured
projected rotational velocity of $v_{\rm B} \sin i = 31$~\kms\ for
that star, we infer that it would have to be seen nearly pole-on
($i_{\rm rot} \sim 7$\arcdeg) in order to be responsible for the
photometric modulation, which we cannot rule out.  For this
calculation, we adopted a radius of $R_{\rm B} = 0.78~R_{\sun}$, based
on the PARSEC isochrone.  The longer rotation periods would lead to
less extreme inclinations relative to the line of sight, of about
14\arcdeg\ and 37\arcdeg, respectively.  As none of these angles agree
with the orbital inclination, they would imply a misalignment between
the spin and orbital axes. Whether or not this is the case, star~B is
itself clearly an UFR, as its measured projected rotational velocity
of 31\thinspace\kms\ implies an upper limit to its rotation period of
1.3~days, much shorter than typical for a star of its spectral type in
the Pleiades.

On the other hand, a possibility that seems more likely to us is that
the rotational modulation signature originates from the primary star,
whose lines we do not detect in our spectra. It is the brighter
component by nearly a factor of two, if we rely on the PARSEC models
for stars of these masses, and very rapid rotation of the primary
would in fact be a natural explanation for its non-detection. In that
case, both components of \hstar\ would fall in the category of UFRs,
and their spin rates would be primordial as tidal forces are
negligibly small in an orbit with such a long period. If the
rotational signal comes from the primary, an assumed radius for the
star of $0.86~R_{\sun}$ would result in equatorial rotational
velocities of about 280, 140, and 55~\kms\ for the three reported
values of $P_{\rm rot}$. The actual line broadening would depend on
the inclination angle of its spin axis projected onto the line of
sight.

\section{Discussion and conclusions}
\label{sec:discussion}

\hstar\ is the fourth system in the Pleiades cluster with dynamical
mass determinations. We have shown that previous claims suggesting it
contains a pair of sharp-lined stars in a presumably short-period
orbit are incorrect, and that based on the currently available
observations, the system is best described as consisting of a
moderately rotating star with visible spectral features in a slightly
eccentric 18-year orbit around a more massive companion.  The masses
we determine from our new spectroscopic and adaptive optics
observations, other radial velocities from the literature, and
previously published VLBI measurements that resolve the pair, have
relative uncertainties of about 2.5\%. They are limited mostly by the
precision of the early radial velocities (pre-1995). The masses
correspond approximately to stars of spectral types G5 and G9. We also
derive a parallax good to better than 0.4\%, which is in excellent
agreement with the (adjusted) value from {\it Gaia}/DR2, though more
precise.

\begin{figure}
\epsscale{1.15}
\plotone{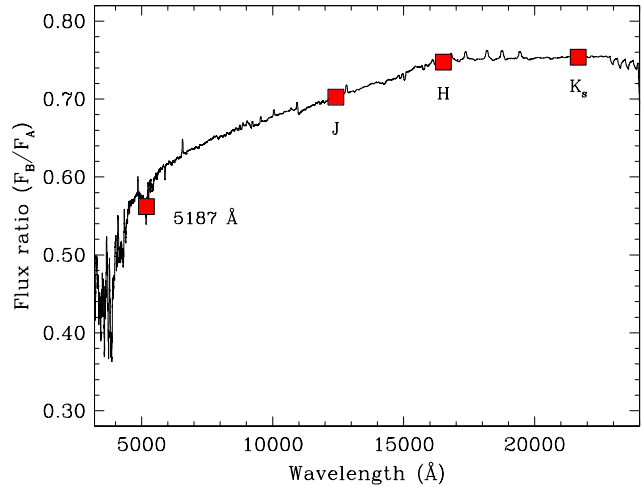}

\figcaption{Flux ratio between stars~B and A as a function of
  wavelength. The curve is based on synthetic spectra by
  \cite{Husser:2013} for solar metallicity and temperatures of
  5750\thinspace K and 5400\thinspace K, normalized using a radius
  ratio of $R_{\rm B}/R_{\rm A} = 0.9$. The values for the 2MASS
  $JHK_{\rm S}$ bandpasses and the wavelength region of our TRES
  spectra ($\sim$5187\thinspace \AA) are indicated with squares, and
  result from integrating over the curve using the corresponding
  transmission functions. \label{fig:fluxratio}}

\end{figure}

The most puzzling aspect of our results is the lack of detection of
the lines of the G5 primary star in our spectra, despite multiple
attempts carried out with TODCOR using a wide range of template
parameters.  According to the PARSEC models, that star is expected to
be roughly 0.6\thinspace mag brighter than the secondary.
Circumstantial evidence that the light of star~A is attenuating the
lines of star~B was presented in Section~\ref{sec:companion}. We now
examine this in a somewhat more quantitative fashion.

As a first step, we made an independent estimate of the brightness
difference between the components using solar-metallicity synthetic
spectra based on PHOENIX model atmospheres from the library of
\cite{Husser:2013}. We adopted effective temperatures of
5750\thinspace K and 5400\thinspace K, near those determined for
stars~A and B, and a radius ratio estimate of $R_{\rm B}/R_{\rm A} =
0.9$ from the PARSEC models. The resulting flux ratio as a function of
wavelength is seen in Figure~\ref{fig:fluxratio}. The values for the
2MASS $JHK_{\rm S}$ bandpasses ($F_{\rm B}/F_{\rm A} = 0.702$, 0.747,
and 0.754, respectively) correspond to magnitude differences of
$\Delta J = 0.38$, $\Delta H = 0.32$, and $\Delta K_{\rm S} = 0.31$,
which are very close to the values measured from our AO observations
(see Table~\ref{tab:ao}). The ratio at the mean wavelength of our
spectroscopic observations (5187\thinspace \AA) is 0.56 ($\Delta m =
0.63$). We therefore expect star~A to be about 1.8 times brighter than
star~B in this spectral region.

\begin{figure}
\epsscale{1.15}
\plotone{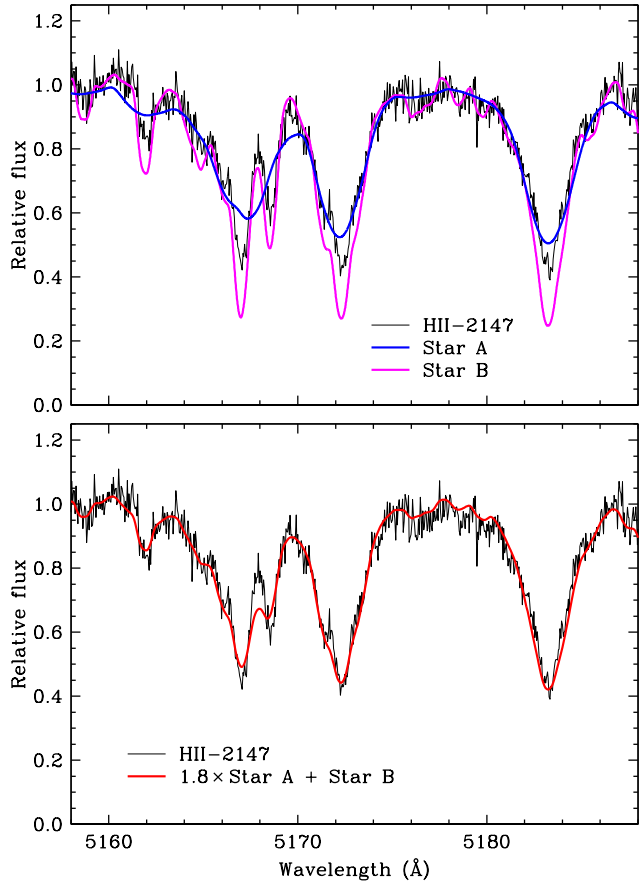}

\figcaption{emph{Top:} Observed spectrum of \hstar\ as in
  Figure~\ref{fig:obsspectra}, along with synthetic spectra to
  represent stars~A ($T_{\rm eff} = 5750$\thinspace K, $v \sin i =
  100$~\kms) and star~B ($T_{\rm eff} = 5500$\thinspace K, $v \sin i =
  30$~\kms). \emph{Bottom:} Comparison between the observed spectrum
  of \hstar\ and the result of creating a synthetic binary spectrum by
  adding 1.8 times the flux of star~A to the flux of star~B (see
  text). \label{fig:synspectra}}

\end{figure}

With this information, we then explored the effect on the strength of
the lines of star~B.  Figure~\ref{fig:synspectra} (top panel)
reproduces the same observed spectrum of \hstar\ from
Figure~\ref{fig:obsspectra}, which was shown there to be affected by
veiling. Also plotted is the synthetic template we used to derive the
radial velocities, which we take to represent star~B without
dilution. This template provides a very good match to the line
profiles of real stars, as illustrated before. To represent star~A we
have chosen a synthetic template from the same library with a
temperature of 5750\thinspace K, and an arbitrary rotational
broadening of 100~\kms.  The bottom panel of
Figure~\ref{fig:synspectra} displays the result of adding together the
flux of star~B and 1.8 times the flux of star~A, after
renormalization. The comparison with the real spectrum of
\hstar\ shows very good agreement in the line depths, providing a
satisfactory explanation of the observed degree of veiling.

We view the result of this exercise as a valuable self-consistency
check on the various aspects of our analysis, including the mass
determinations. It also supports the notion advanced earlier in the
paper that rapid rotation is, in fact, responsible for the
non-detection of the lines of star~A in our spectra.

Although {\it Gaia\/} will not spatially resolve the 18-yr pair and
can only measure the motion of its center of light, by the end of the
mission it is possible that it could provide some constraints on all
of the elements of the A-B orbit (which has the same shape as the
photocenter orbit), except for the semimajor axis.  Direct
spectroscopic detection of star~A may be possible with observations of
higher signal-to-noise ratio than we have available. In that case,
measurement of its radial velocities would result in a better
constrained semiamplitude $K_{\rm A}$, which would in turn improve the
mass determinations further.


\begin{acknowledgements}

The spectroscopic observations of \hstar\ were gathered with the help
of P.\ Berlind, M.\ Calkins, and G.\ Esquerdo. We thank D.\ Latham for
organizing those observations, and J.\ Mink for maintaining the CfA
echelle database. We also thank the anonymous referee for helpful
comments.  G.T.\ acknowledges partial support from the National
Science Foundation (NSF) through award AST-1509375.
C.M.\ acknowledges support from the NSF under award AST-1313428.
Research at Lick Observatory is partially supported by a generous gift
from Google. Some of the data presented herein were obtained at the
W.M.\ Keck Observatory, which is operated as a scientific partnership
among the California Institute of Technology, the University of
California and the National Aeronautics and Space Administration. The
Observatory was made possible by the generous financial support of the
W.M.\ Keck Foundation.  The research has made use of the SIMBAD and
VizieR databases, operated at the CDS, Strasbourg, France, of NASA's
Astrophysics Data System Abstract Service, and of the WEBDA database,
operated at the Department of Theoretical Physics and Astrophysics of
the Masaryk University (Czech Republic). The work has also made use of
data from the European Space Agency (ESA) mission {\it Gaia}
(\url{https://www.cosmos.esa.int/gaia}), processed by the {\it Gaia}
Data Processing and Analysis Consortium (DPAC,
\url{https://www.cosmos.esa.int/web/gaia/dpac/consortium}). Funding
for the DPAC has been provided by national institutions, in particular
the institutions participating in the {\it Gaia} Multilateral
Agreement. The computational resources used for this research include
the Smithsonian Institution's ``Hydra'' High Performance Cluster
(\url{https://doi.org/10.25572/SIHPC}).

\end{acknowledgements}

\appendix

We describe here our procedure to develop a relation to predict the
magnitudes of Pleiades stars in the {\it Gaia\/} and 2MASS bandpasses
as a function of mass, which we used in Section~\ref{sec:analysis} to
verify the accuracy of Solution~1 and to explore alternate
configurations for \hstar\ involving three stars.

We began by establishing purely empirical relations between color and
brightness using the list of 1454 likely Pleiades members published
recently by \cite{Gao:2019}, based on astrometric and photometric
observations from the {\it Gaia}/DR2 catalog. We fitted cubic spline
relations to the (single-star) main sequence in diagrams of absolute
$G$, $G_{\rm BP}$, and $G_{\rm RP}$ magnitude as a function of the
observed $G_{\rm BP}-G_{\rm RP}$ color, where the absolute magnitudes
were computed using the parallax of each star to reduce scatter due to
the non-negligible depth of the cluster. Note that all of these
photometric quantities are affected by extinction, although the effect
is relatively small in the Pleiades \citep[$E(B-V) \approx
  0.04$~mag;][]{Taylor:2008}.\footnote{Reddening is not uniform across
  the Pleiades cluster, being smaller on the eastern side than the
  western side \citep[see][]{Breger:1986, Taylor:2008}. As individual
  estimates are not available for all 1454 members, we have chosen to
  adopt here an average $E(B-V)$ of 0.04 that suffices for this work.}
Next, to provide the necessary connection with stellar mass, we
adopted a mapping between mass and theoretical $G_{\rm BP}-G_{\rm RP}$
color from the 125~Myr, solar metallicity PARSEC isochrone used in the
main text. The model colors were adjusted by applying reddening in the
amount of $E(G_{\rm BP}-G_{\rm RP}) = 1.31 E(B-V)$
\citep{Stassun:2019}, to make them more consistent with the observed
colors (i.e., with the abscissa of the empirical spline relations).
Reliance on models for the mass-color mapping will of course not be
perfect because of deficiencies in the model fluxes (due, e.g., to
missing opacity sources), but is unavoidable, and in the end this
two-step procedure allows us to predict the brightness and color of a
star from its mass more accurately than using the models alone. We
illustrate this below.  In a similar fashion, we developed spline
relations to predict the $J$, $H$, and $K_{\rm S}$ magnitudes in the
2MASS system as a function of the observed $G_{\rm BP}-G_{\rm RP}$
color.

\begin{figure}[!ht]
\epsscale{0.9}
\plotone{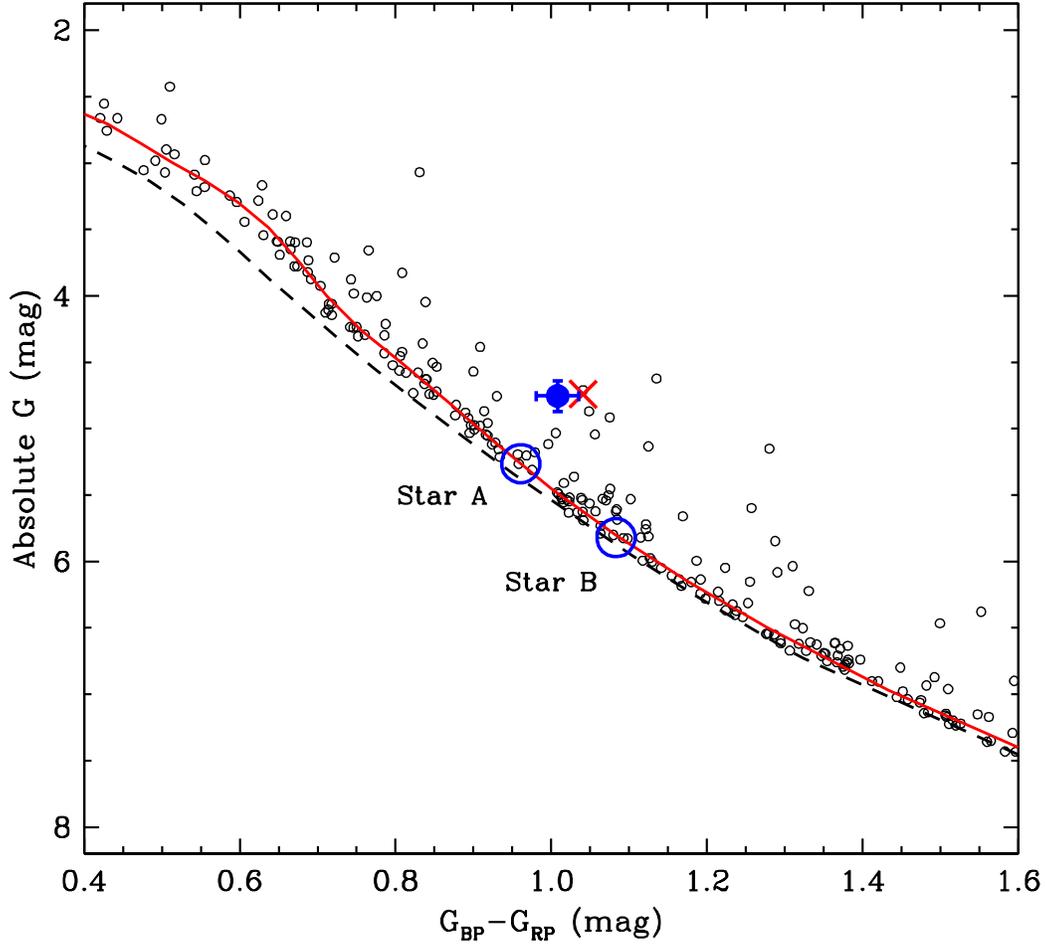}

\figcaption{Color-magnitude diagram for likely Pleiades members from
  \cite{Gao:2019}. Absolute magnitudes were derived using the
  individual parallax of each star. The location of \hstar\ above the
  single-star sequence, as measured by {\it Gaia}, is indicated by the
  red cross. The solid red line is based on our semi-empirical
  mass-magnitude relations, and the blue open circles on it mark the
  predicted location of the binary components according to their
  masses of 0.978 and $0.897~M_{\sun}$, as given in
  Table~\ref{tab:mcmc} (Solution~2). The blue dot with error bars
  represents the combined light of the binary inferred using the
  semi-empirical relations, with the errors being dominated by the
  uncertainty in the masses (photometric errors are much smaller). The
  theoretical color-magnitude relation as given by the PARSEC
  isochrone mentioned in the text (dashed line) shows a poorer fit to
  the observations. All magnitudes and colors shown are affected by
  extinction and reddening. \label{fig:deconvolve}}

\end{figure}

Figure~\ref{fig:deconvolve} shows the location of the two stars in the
color-magnitude diagram of the Pleiades based on {\it Gaia\/}
photometry, together with other cluster members from the list of
\cite{Gao:2019}. The observed location of \hstar\ above the
single-star sequence is represented by the red cross, and the point
with error bars marks the predicted location of the combined light
according to the masses derived in Solution~2, based on our
semiempirical spline relations (red solid line). The dashed line
represents the 125~Myr solar-metallicity PARSEC isochrone for the
Pleiades \citep{Chen:2014}, which does not provide as good a match to
the observations.

As a sanity check, we tested the ability of these relations to predict
the true colors of stars as a function of mass by using the few
examples in the Pleiades that are in binary systems and have
dynamically measured masses. One of them, Atlas, has its brightness
measurements compromised by saturation, and is also beyond the range
of our calibrations. The other two, HD~23642 and HCG~76
\citep{David:2016}, are within the validity range, but are near the
edges of our relations (upper and lower ends, respectively).
Nevertheless, with the individual masses in each of these systems, we
predicted their brightness in the {\it Gaia\/} bandpasses and then
calculated the $G_{\rm BP}-G_{\rm RP}$ color index for the combined
light, for comparison with the measurements from {\it Gaia\/}. For
HD~23642 our relations predict a $G_{\rm BP}-G_{\rm RP}$ index of
0.096~mag, whereas the measured value is 0.107~mag. For HCG~76 we
obtain $G_{\rm BP}-G_{\rm RP} = 2.925$~mag, and the measured value is
2.913~mag. Given that the reddening toward these two objects may be
different than the mean value we have adopted for the cluster, we
consider these differences ($-0.011$ and +0.012~mag) to be small, and
therefore to support the accuracy or our calibration.

\end{document}